\documentclass[12pt,tightenlines,eqsecnum,floats,showpacs,nofootinbib,amsmath,amssymb,aps,prd]{revtex4}
\usepackage{graphicx,verbatim}
\usepackage{amsmath}
\usepackage{amsfonts}
\usepackage{amssymb}
\usepackage[colorinlistoftodos]{todonotes}
\usepackage{epstopdf}

\def\grad{\nabla}
\def\Lie{\mathcal{L}}
\def\scri{\mathcal{I}}
\def\rmd{\mathrm{d}}
\def\={\hat{=}}
\def\L{\mathcal{L}}
\def\S{\mathcal{S}}
\def\T{\mathcal{T}}
\def\scri{\mathcal{I}}

\def\F{\mathcal{F}}

\newcommand{\pb}[1]{\hbox{\lower0.5ex\hbox{${}_{\leftarrow}$}}\kern-1.9ex{#1}}

\def\Gds{G_{\rm dS}}
\def\G{\mathfrak{G}}
\def\Diff{{\rm Diff}(\mathcal{I})}
\def\B{\mathfrak{B}}
\def\LBMS{\mathfrak{b}}
\def\LS{\mathfrak{s}}
\def\g{\mathfrak{g}}

\def\ub{\underbar}

\def\be{\begin{equation}}
\def\ee{\end{equation}}
\def\ba{\begin{eqnarray}}
\def\ea{\end{eqnarray}}

\def\f{\frac}

\def\qo{\mathring{q}}
\def\Co{\mathring{c}}
\def\Ko{\mathring{k}}
\def\To{\mathring{t}}
\def\eo{\mathring{e}}
\def\Do{\mathring{D}}
\def\no{\mathring{n}}

\def\SO(3){\rm SO(3)}
\def\so(3){\rm so(3)}
\def\SO(4){\rm SO(4)}
\def\so(4){\rm so(4)}
\def\SO(1,4){\rm SO(1,4)}
\def\so(1,4){\rm so(1,4)}
\def\SU(2){\rm SU(2)}

\begin{document}

\title{Asymptotics with a positive cosmological constant:\\ I. Basic framework} 
\author{Abhay Ashtekar}
\email{ashtekar@gravity.psu.edu} 
\author{B\'eatrice Bonga}
\email{bpb165@psu.edu} 
\author{Aruna Kesavan}
\email{aok5232@psu.edu} \affiliation{Institute for Gravitation and the
Cosmos \& Physics Department, Penn State, University Park, PA 16802,
U.S.A.}

\begin{abstract}
The asymptotic structure of the gravitational field of isolated systems  has been analyzed in great detail in the case when the cosmological constant $\Lambda$ is zero. The resulting framework lies at the foundation of research in diverse areas in gravitational science. Examples include: i) positive energy theorems in geometric analysis; ii) the coordinate invariant characterization of gravitational waves in full, non-linear general relativity; iii) computations of the energy-momentum emission in gravitational collapse and binary mergers in numerical relativity and relativistic astrophysics; and iv) constructions of asymptotic Hilbert spaces to calculate $S$-matrices and analyze the issue of information loss in the quantum evaporation of black holes. However, by now observations have established that $\Lambda$ is positive in our universe. In this paper we show that, unfortunately, the standard framework does not extend from the $\Lambda =0$ case to the $\Lambda >0$ case in a physically useful manner. In particular, we do not have positive energy theorems, nor an invariant notion of gravitational waves in the non-linear regime, nor asymptotic Hilbert spaces in dynamical situations of semi-classical gravity. A suitable framework to address these conceptual issues of direct physical importance is developed in subsequent papers.
\end{abstract}

\pacs{04.70.Bw, 04.25.dg, 04.20.Cv}
\maketitle

\section{Introduction}
\label{s1}

The analysis of asymptotic structure of the gravitational field is rather subtle in general relativity because the field of interest itself determines the space-time geometry that is needed to specify the boundary conditions at infinity. The ensuing difficulties have been systematically overcome starting with the pioneering work of Arnowitt, Deser and Misner (ADM) at spatial infinity \cite{adm,aarh,spatial-review} and Bondi, Sachs, Newman and Penrose at null infinity \cite{bondi,null-review}. The resulting frameworks continue to lie at the foundation for a large body of research in classical and quantum gravity.

The ADM framework provided an invariant notion of the total energy-momentum of isolated gravitating systems. Because the gravitational potential energy is negative, at first it was not clear whether the total energy is always positive. In fact, in the 1970s serious attempts were made to construct counter-examples. However, subsequent research established that the ADM energy is necessarily positive \cite{sy,ew} and that the total 4-momentum is necessarily time-like \cite{aagh} so long as matter satisfies suitable energy conditions. The ADM energy thus provides a new invariant for certain asymptotically flat 3-manifolds which has sparked significant research at the interface of general relativity and geometric analysis. 

Applications of the Bondi framework \cite{bondi,null-review} span an impressive array of contemporary research. Prior to the introduction of this framework, there was considerable controversy on the physical reality of gravitational waves. Already in 1917, Einstein had isolated the radiative modes of the gravitational field \emph{in the linear approximation} and derived the celebrated quadrupole formula. The debate was whether this was an artifact of linearization; i.e., if one could really distinguish \emph{physical} gravitational waves from coordinate effects in \emph{full, non-linear} general relativity.%
\footnote{This discussion was not limited just to a fringe of the research community. Surprising as it may now seem, leading thinkers including Einstein and Eddington argued against the physical reality of gravitational waves. See e.g. \cite{history}.} 
The C-metric, discovered by Levi-Civita \cite{lc} in 1919 provides a good illustration. This solution to Einstein's equations appeared to admit gravitational waves but it also appeared to be stationary, prompting many researchers to believe that the waves were merely coordinate artifacts. It was much later that a detailed analysis of this metric using null infinity firmly established that the `pair of black holes' described by the metric does emit energy in the form of gravitational waves \cite{bicak1,aatd}; there is no contradiction because what was believed to be a `stationary' Killing vector is in fact a boost \cite{kw}. 

Bondi and his coworkers resolved the initial confusion by constructing a framework in which the issue could be analyzed in an invariant fashion. Specifically, because gravitational waves propagate along null cones in general relativity, they constructed a systematic expansion of the metric as one moves away from the sources in \emph{null} directions and studied asymptotics at null infinity, in contradistinction to the ADM framework which focuses on spatial infinity. This construction was cast in a more convenient form through conformal techniques by Penrose \cite{rp} where null infinity is represented as the boundary, $\scri$, of the physical space-time in its conformal completion. From the curvature of the gravitational connection $D$ defined intrinsically on $\scri$, one can then construct an invariant, second rank, trace-free, transverse tensor field $N_{ab}$ on $\scri$, called the \emph{Bondi news} \cite{aa-asym}. The two free components of $N_{ab}$ provide an invariant characterization of the two radiative modes of the gravitational field in full general relativity. Thus a given asymptotically flat space-time admits gravitational radiation if and only if $N_{ab} \not=0$. Fluxes of energy and momentum emitted by an isolated gravitating system, such as a compact binary, are expressed as integrals of $|N_{ab}|^{2}$, with suitable weights that correspond to the component of energy-momentum of interest \cite{aams}. (In particular, $N_{ab}\not=0$ for the Levi-Civita C-metric \cite{aatd}). Finally, the non-triviality of the distinction between exact general relativity and its linearized approximation is brought to the forefront by the fact that, in presence of gravitational radiation, \emph{even near $\scri$, one cannot tease out a canonical flat metric $\eta_{ab}$ from the physical metric $g_{ab}$} to obtain an unambiguous expansion $g_{ab} = \eta_{ab} + (1/r) h_{ab}$ to this leading order. As a result, the asymptotic symmetry group is \emph{not} the Poincar\'e group as one would have first expected, but an infinite dimensional generalization thereof, the Bondi Metzner Sachs (BMS) group $\B$ \cite{bondi,aa-asym}. Consequently, if $N_{ab}\not=0$, there is an infinite dimensional, `super-translation' ambiguity in the definition of angular momentum of isolated gravitational systems in general relativity \cite{aams,td}. In quantum theory, this enlargement is directly related to the infrared issues associated with the full, non-linear gravitational field \cite{aa-asym,as}.  

The Bondi et al framework provides the conceptual basis for calculations of gravitational radiation in numerical simulations and in arriving at their astrophysical implications. The waveforms that play a key role at the interface of numerical relativity and data analysis refer to the connection $D$, or its curvature $N_{ab}$, or the time derivative of $N_{ab}$, encoded in the component $\Psi_{4}$ of the asymptotic Weyl tensor at $\scri$. The BMS group $\B$ admits a unique 4-dimensional, Abelian normal sub-group $\T$ which, in Minkowski space, can be naturally identified with the group of translations \cite{sachs}. One uses this group to calculate the fluxes of energy-momentum across $\scri$. For example, in a binary coalescence, one finds that the gravitational radiation carries away 3-momentum in the center of mass frame, whence the final black hole receives a `kick' in this frame \cite{kicks} with interesting implications to astrophysics of both solar-mass and supermassive black holes. Had the asymptotic symmetry group been just the full diffeomorphism group $\Diff$, these calculations would have no coordinate invariant significance. Thus, while the Bondi framework is not always explicit in these calculations, it lies at their foundation. Conceptual subtleties such as the structure of the BMS group $\B$ and the precise meaning of gravitational radiation have a direct impact on the final physical results.

Penrose's null infinity $\scri$ and the translation sub-group $\T$ of the BMS group $\B$ also lie at the heart of the construction of asymptotic Hilbert spaces in quantum field theory of zero rest-mass fields in black hole space-times. For, it is the availability of $\T$ that enables one to decompose fields into positive and negative frequency parts on $\scri$ ---or, to select uniquely the complex structures on the phase spaces of these fields--- in order to construct the Fock representations. One then finds that the one particle Hilbert spaces provide unitary, irreducible representations of all of the Poincar\'e sub-groups of $\B$. The corresponding Casimir operators then unambiguously attribute zero mass and appropriate spin to these fields \cite{aa-asym}. Note that these Hilbert spaces are well-defined even when the underlying space-time is non-stationary; $N_{ab}$ need not vanish at $\scri$. In particular then, these constructions are meaningful on the dynamical semi-classical space-times that include the back reaction from quantum evaporation of black holes \cite{aafpfr}. Had this not been true, then there would be no adequate asymptotic Hilbert spaces to define a $S$-matrix and the question of its unitarity and information loss could not even be phrased. Finally, thanks to this structure at $\scri$, one can construct asymptotic Hilbert spaces even for the radiative modes of the full, non-linear gravitational field \cite{aa-asym} and analyze conceptual issues such as whether CPT will be violated in full quantum gravity.

All this rich structure refers to the case $\Lambda=0$. What is the situation for $\Lambda > 0$? Penrose's construction of null infinity naturally generalizes; he showed this already in his first papers. $\scri$ is again a boundary of the physical space-time within its conformal completion. \emph{But it is now space-like.} Consequently, as we will see in detail, the asymptotic symmetry group ---the direct analog of the BMS group $\B$--- is now the the full diffeomorphism group $\Diff$ of $\scri$. \emph{There is no natural analog of the Bondi news to characterize gravitational radiation in the full non-linear context.} Indeed, in the description of isolated gravitating systems, there is no known strategy to impose even the elementary requirement of `no incoming radiation condition' at $\scri^{-}$.  

A common strategy is to try to improve on this situation by imposing an extra requirement: demand that the intrinsic geometry of $\scri$ be conformally flat. At first, this strengthening seems very natural because in the conformal completion of de Sitter space-time (that replaces Minkowski space-time in the transition from $\Lambda=0$ to $\Lambda >0$) the intrinsic geometry of $\scri$ \emph{is} in fact conformally flat. Furthermore, once this condition is imposed, as we will see, the asymptotic symmetry group reduces from $\Diff$ precisely to the 10-dimensional de Sitter group $\Gds$! Therefore, one can now hope to define 10 de Sitter charges ---that would naturally extend the notion of the Bondi momentum--- and fluxes of these charges across $\scri$ would provide us with expressions analogous to the fluxes of Bondi momentum, now for the $\Lambda >0$ case. Therefore the strategy seems very attractive. Additional support for it comes from the much studied $\Lambda <0$ case, where one does generally ask that the intrinsic metric of $\scri$ be conformally flat \cite{aaam,ht}. For, in the  $\Lambda<0$ case, $\scri$ is \emph{time-like} whence, in any case, one needs an additional boundary condition to make the evolution well-defined and it turns out that the conformal flatness requirement can be regarded as a natural `reflective boundary condition' \cite{swh,aaam}. 

But for $\Lambda >0$, as we just pointed out, $\scri$ is space-like, and a `reflective boundary condition', or indeed \emph{any} boundary condition would be an additional restriction on the space of permissible initial data at $\scri$. One might hope that these extra conditions are somehow harmless. But a detailed examination will show that this is unfortunately \emph{not} the case! It is a genuine restriction \emph{that removes by hand half the permissible data} and this elimination has no physical basis whatsoever. In fact the restriction is so severe that it rules out non-zero fluxes of de Sitter charges. Thus, if we imposed this condition we would find that \emph{gravitational waves do not carry away (de Sitter) energy or momentum across $\scri$!} We will show in detail how these consequences arise. But if we do not impose this additional boundary condition, we seem to face the opposite difficulty. Now the structure at $\scri$ is so weak that none of the advances in the $\Lambda=0$ case we discussed above will extend to the $\Lambda>0$ case, \emph{no matter how small $\Lambda$ is.} The goal of this series of papers is to show that one can supplement this framework appropriately so as to address issues of direct physical interest in the $\Lambda >0$ universe we inhabit. A brief summary of this program can be found in \cite{abk-letter}.

In this first detailed paper, we focus on the standard strategies and show how the difficulties discussed above arise. The paper is organized as follows. In section \ref{s2} we summarize the notion of asymptotically de Sitter space-times. The basic definition due to Penrose is extended to incorporate isolated systems including black holes and the cosmological space-times as they are commonly treated in the literature. We also discuss the asymptotic fields and equations they satisfy at $\scri$. All this structure will be useful not only in the sections that follow but also in subsequent papers. Section \ref{s3} discusses examples. 
The Vaidya-de Sitter solution is particularly interesting because it brings out certain features associated with non-trivial dynamics. Section \ref{s4} discusses symmetries and the associated definitions of conserved charges. In particular, the role played by the additional requirement of conformal flatness of the intrinsic metric on $\scri$ is spelled out. At first this framework seems satisfactory. Section \ref{s5} explains in detail why this is not the case. Section \ref{s6} summarizes the results and discusses other conceptual issues that arise in the passage from $\Lambda=0$ to $\Lambda >0$ cases. These can be addressed by the new strategy introduced in the second \cite{abk2} and the third \cite{aads} papers in this series. 

Our conventions are as follows. Throughout we assume that the underlying space-time is 4-dimensional and the space-time metric has signature -,+,+,+. Physical fields will carry hats while those which are well defined on the conformal completion will be unhatted. The curvature tensors (in the completion) are defined via:  $2\nabla_{[a}\nabla_{b]} k_c =
R_{abc}{}^d k_d$, $R_{ac} = R_{abc}{}^b$ and $R = R_{ab}g^{ab}$.

\section{Asymptotically de Sitter space-times}
\label{s2}
This section is divided in two parts. In the first, we present definitions of asymptotically de Sitter space-times by mimicking the procedure used in asymptotically anti-de Sitter space-times \cite{aaam,aasd} (see also \cite{ht}). While the basic underlying idea is completely parallel to that used in the $\Lambda=0$ case, it is now natural to allow for three different topologies of $\scri$ which arise in the most common applications. 
In the second part, we summarize the basic consequences of the conditions in the definition. These results will be used in sections \ref{s4}-\ref{s5}.

\subsection{Definitions}
\label{s2.1}

\textbf{Definition 1:} A space-time $(\hat{M},\hat{g}_{ab})$ will be said to be \textit{weakly asymptotically de Sitter} if there exists a manifold $M$ with boundary $\scri$ equipped with a metric $g_{ab}$ and a diffeomorphism from $\hat{M}$ onto $(M\, \setminus\, \scri)$ of $M$ (using which we identify $\hat{M}$ and ($M\, \setminus\, \scri$)) and the interior of $M$ such that: 

i)\, there exists a smooth function $\Omega$ on $M$ such that $g_{ab}=\Omega^2 \hat{g}_{ab}$ on $\hat{M}$; $\Omega=0$ on $\scri$;\\ 
\indent\indent and $n_a := \grad_a \Omega$ is nowhere vanishing on $\scri$;\, and

ii)\,$\hat{g}_{ab}$ satisfies Einstein's equations with a positive cosmological constant,\\ \indent\indent i.e., $\hat{R}_{ab} - \frac{1}{2} \hat{R} \hat{g}_{ab} + \Lambda \hat{g}_{ab} = 8 \pi G \; \hat{T}_{ab}$ with $\Lambda >0$; where $\Omega^{-1} \hat{T}_{ab}$ has a smooth limit to $\scri$.\\

The two conditions in this definitions are direct generalizations of those used in the asymptotically Minkowski space-times \cite{asymflat}. The first ensures that $(M, g_{ab})$ is a conformal completion of the physical space-time $(\hat{M},\, \hat{g}_{ab})$ in which the boundary $\scri$ is at infinity with respect to the physical metric $\hat{g}_{ab}$. The condition $\nabla_{a}\Omega \not=0$ ensures that $\Omega$ can be used as a coordinate on $M$; we can perform Taylor expansions in $\Omega$ to capture the degree of fall-off of physical fields. In terms of the physical space-time $(\hat{M},\, \hat{g}_{ab})$, it ensures that $\Omega$ `falls-off as $1/r$'\, i.e., has the same asymptotic behavior as in de Sitter space. The second condition ensures that the matter fields fall-off appropriately in the physical space-time $(\hat{M},\, \hat{g}_{ab})$. The specific fall-off of $\hat{T}_{ab}$ used here is motivated by the analysis of test fields in de Sitter space-times. Standard matter fields such as the conformally coupled scalar field and the Maxwell field, as well as more phenomenological matter fields such as dust and fluids satisfy this condition. (In fact null fluids and Maxwell fields falls-off faster; $\Omega^{-2} \hat{T}_{ab}$ has a smooth limit to $\scri$.)

\emph{Remarks}:\\ \indent
1) There is considerable freedom in the choice of the conformal factor. Given an admissible conformal factor $\Omega$,\,\, $\Omega^{\prime}=\omega \Omega$ is also admissible provided $\omega$ is smooth and nowhere zero on $M$. One must make sure that physical quantities constructed from the rescaled metric $g_{ab}$ are all invariant under this rescaling. 

2) We will see below that, because $\Lambda >0$ (rather than $\Lambda =0$), $\scri$ is necessarily space-like (rather than null). Still, as in asymptotically Minkowski space-times, we can have both future and past boundaries $\scri^{+}$ and $\scri^{-}$: The causal future of $\scri^{+}$ is $\scri^{+}$ itself and the causal past of $\scri^{-}$ is $\scri^{-}$ itself.  In what follows, unless otherwise stated, \emph{by $\scri$ we will mean $\scri^{+}$ \emph{or} $\scri^{-}$.}\\

For certain global issues ---in particular for a satisfactory definition of a black hole--- one needs to strengthen Definition 1 by 
requiring that $\scri$ be complete.\\
\textbf{Definition 2:} A weakly asymptotically de Sitter space-time $(\hat{M},\hat{g}_{ab})$ will be said to be \textit{asymptotically de Sitter} if $\scri$ is geodesically complete w.r.t. $g_{ab}$.\\

So far we did not specify the topology of $\scri$. Three topologies are of special interest for the most important of physical applications. Therefore we introduce the following classification of 
asymptotically de Sitter space-times.
\begin{itemize}
\item $(\hat{M}, \hat{g}_{ab})$ will be said to be \textit{Globally asymptotically de Sitter} if it admits a conformal completion satisfying conditions of Definition 1 in which $\scri$ has the topology of a 3-sphere $\mathbb{S}^3$. de Sitter space-time with its standard completion belongs to this class.\\
\item $(\hat{M}, \hat{g}_{ab})$ will be said to be \textit{Asymptotically de Sitter in a Poincar\'e patch} if it admits a conformal completion satisfying conditions of Definition 1 in which its $\scri$ has topology $\mathbb{R}^3 \simeq \mathbb{S}^3\, \setminus\, \{p\}$. The standard completions of the Friedmann-Lema\^itre cosmologies, for example, belong to this class where the point $p$ represents spatial infinity, $i^{o}$. Therefore, this topology is of interest particularly in cosmological applications.\\
\item $(\hat{M}, \hat{g}_{ab})$ will be said to be \textit{Asymptotically Schwarzschild-de Sitter} if it admits a conformal completion satisfying conditions of Definition 1 in which its $\scri$ has topology $\mathbb{S}^2 \times \mathbb{R} \simeq \mathbb{S}^3\, \setminus\, \{p_{1}, p_{2}\}$ on $\scri$. The standard completion of Schwarzschild-de Sitter space-time falls in this class, where the point $p_{2}$ again represents spatial infinity $i^{o}$  and the point $p_{1}$ represents future or past time-like infinity $i^{\pm}$ on $\scri^{\pm}$. This topology is of interest in the discussion of compact isolated systems such as stars and black holes.
\end{itemize}

Note that a physical space-time can belong to more than one class, depending on the choice of the conformal factor. For example, given the standard conformal completion of de Sitter space-time in which $\scri$ has $\mathbb{S}^{3}$ topology, one can choose another conformal factor $\omega^{\prime} = \alpha\Omega$ to obtain a completion in which it has $\mathbb{R}^{3}$ topology: the required $\alpha$ will simply diverge at a point $p$ on $\mathbb{S}^{3}$ at an appropriate rate, `opening up' $\mathbb{S}^{3}$, and the point $p$ would then have the interpretation of being the point $i^{o}$ at spatial infinity. The new completion would make $(M,g_{ab})$ `asymptotically de Sitter in a Poincar\'e patch'. Similarly, given the standard $\mathbb{S}^{3}$ completion, we can choose a conformal factor $\Omega^{\prime} = \beta\Omega$ which diverges at appropriate rates at two \emph{points} which will represent $i^{o}$ and $i^{\pm}$ on $\scri^{\pm}$ of the resulting completion. 

However, given the standard completion of the Schwarzschild-de Sitter space-time in which $\scri$ is topologically $\mathbb{S}^{2}\times \mathbb{R}$, it is not possible to choose a conformal rescaling $\beta$ to obtain a smooth rescaled metric $g_{ab}$ and a $\mathbb{S}^{3}$ topology for $\scri$. The detailed discussion contained in the subsequent sections will make these features transparent. Here we only note that in the asymptotically Minkowski context, the topology of $\scri$ is always $\mathbb{S}^{2}\times \mathbb{R}$, irrespective of whether the physical space-time of interest is just the Minkowski space-time or represents a star or a black hole. This difference arises because, as we will discuss in sections \ref{s3} and \ref{s4}, whereas $\scri$ is naturally ruled (by null geodesics) if $\Lambda=0$, there is no such ruling in the case when $\Lambda>0$. One can also consider a topology $\mathbb{S}^{3}\, \setminus\, \{p_{1},p_{2}, \ldots p_{n}\}$. In this case, one of the punctures will represent the point $i^{o}$ at spatial infinity and the remaining $n-1$ punctures would represent compact objects. However, since these are distinct points on $\scri^{\pm}$, the physical distance between any two of them grows unboundedly in the distant future/past. Therefore these cases will not be relevant to the study of individual isolated systems normally considered in mathematical and numerical general relativity. 

Finally, as discussed in section \ref{s1}, sometimes a stronger restriction is imposed to reduce the asymptotic symmetry group and define conserved gravitational charges in the $\Lambda \not=0$ case. This additional boundary condition leads us to the final definition.\\
\textbf{Definition 3:}  A space-time $(\hat{M}, \hat{g}_{ab})$ will be said to be \textit{strongly asymptotically de Sitter} if in a conformal completion satisfying conditions of Definition 2, the intrinsic metric $q_{ab}$ on $\scri$ is conformally flat.\\

As discussed in section \ref{s2.2} below, this condition turns out to be equivalent to requiring that the leading order piece of the Weyl tensor have no magnetic part at $\scri$ \cite{aaam}. This stronger condition is satisfied in the simplest examples normally considered ---de Sitter, Friedmann-Lema\^itre and Kerr-de Sitter--- as well as Vaidya-de Sitter. However, we will see in \cite{abk2} that the condition fails (to the appropriate leading order) already when one allows perturbations representing linearized gravitational waves on these space-times.

\subsection{Asymptotic fields and their equations}
\label{s2.2}

In this sub-section, we will collect the immediate implications of Definition 1. The reasoning used is completely parallel to that in the $\Lambda <0$ case \cite{aaam,aasd}. However, since not all readers will be familiar with the details of the asymptotically anti-de Sitter space-times, and since these results play an important role in the subsequent analysis, we will summarize the main steps. In this discussion, the requirement of completeness in Definition 2 and the choice of topology will not play any role as the considerations of this sub-section are local to $\scri$. 

Let us begin by expressing Einstein's equation satisfied by $\hat{g}_{ab}$ in terms of the conformally rescaled metric $g_{ab}$:
\begin{align}
R_{ab} - \frac{1}{2} g_{ab} R + 2 \Omega^{-1} \left(\grad_a n_b - g_{ab} \grad^c n_c \right) + 3 \Omega^{-2} g_{ab} n^c \, n_c + \Omega^{-2} \Lambda g_{ab} = 8 \pi G \hat{T}_{ab}\, , \label{eqeinstein}
\end{align}
where, as before $n_{a} := \nabla_{a} \Omega$. Multiplying (\ref{eqeinstein}) by $\Omega^2$ and evaluating the resulting expression on $\scri$ using our boundary conditions in Definition 1, we obtain
\begin{align}
n^{a} n_{a}\, \hat{=} \, - \frac{\Lambda}{3} =: - \frac{1}{l^{2}}.
\end{align}
\emph{Here and throughout the paper $\hat{=}$ stands for equality at} $\scri$ and $\l$ denotes the cosmological radius. Thus, $n^a$ is time-like on $\scri$ and consequently $\scri$ is space-like. Hence condition iii) of geodesic completeness in Definition 2  is equivalent to completeness with respect to the Riemannian metric $q_{ab}$ on $\scri$ induced by $g_{ab}$. The space-like character of $\scri$ immediately gives rise to some conceptual complications. For example the `obvious' strategy to impose the no incoming boundary condition fails already in the case of Maxwell fields and one cannot repeat the proofs of `peeling theorems' that were used heavily in the early stages of the analysis of gravitational radiation in the Bondi et al program (see, e.g., \cite{bicaketal,penrose}). 

Next, recall that there is considerable freedom in the choice of $\Omega$. It is easy to verify that one can use this freedom to go to a conformal frame in which $\grad_a n^a \hat{=}0$. Throughout our analysis we will make this restriction because this choice simplifies calculations considerably \cite{aasd}. In particular, Eq. (\ref{eqeinstein}) now implies that $\grad_a n_b \hat{=}0$, whence, in particular, the extrinsic curvature, $k_{ab}$ of $\scri$ vanishes. Note, however, that our restriction on $\Omega$ still leaves a residual conformal freedom: $\Omega \to \Omega^{\prime}= \omega\Omega$ where the derivative of $\omega$ orthogonal to $\scri$ vanishes, $n^a \grad_a \omega \hat{=}0$.

Using this conformal frame we will now show that the Weyl tensor $C_{abcd}$ of $g_{ab}$ vanishes identically on $\scri$. Recall first that the Schouten tensor $S_{ab} := R_{ab} - ({R}/{6})\, g_{ab}$ of $g_{ab}$ is related to that of $\hat{g}_{ab}$ via
\begin{align} \label{S}
\hat{S}_{ab} = S_{ab} + 2 \Omega^{-1} \grad_a n_b - \Omega^{-2} g_{ab}\, n^c n_c,
\end{align}
and that the Riemann tensor can be expressed as
\be
R_{abcd} = C_{abcd} + g_{a[c} S_{d]b} - g_{b[c} S_{d]a}\, , \label{eqriemannweyl}\ee
where $C_{abcd}$ is the Weyl tensor. Taking the `curl' of (\ref{S}) and using (\ref{eqriemannweyl}), one obtains
\be
\grad_{[a} (\Omega \hat{S}_{b]c}) = \Omega \grad_{[a} S_{b]c} + C_{abcd} n^d + g_{c[a} \hat{S}_{b]d} n^d\, , \label{eqcurls}
\ee
and Einstein's equations in the physical space-time imply
\be
\hat{S}_{ab} = \frac{\Lambda}{3} \hat{g}_{ab} + 8\pi G \left( \hat{T}_{ab} - \frac{1}{3} \hat{T} \hat{g}_{ab} \right) \,\equiv\, \frac{\Lambda}{3} \hat{g}_{ab}  + \tilde{T}_{ab} \, \label{eqsrelatedtot}
\ee
where $\hat{T} = \hat{T}_{pq} \hat{g}^{pq}$.
Next, substituting \eqref{eqsrelatedtot} in \eqref{eqcurls} one obtains
\be
\Omega \grad_{[a} S_{b]c} + C_{abcd} n^d  = \grad_{[a} (\Omega \tilde{T}_{b]c} ) - g_{c[a} \tilde{T}_{b]d}\, n^d \label{eqsandweyl}
\ee
Since $\Omega^{-1} \hat{T}_{ab}$ has a smooth limit to $\scri$, we conclude
\be
C_{abcd}\, n^d \,\hat{=} \, 0 \label{eqcdotniszero}
\ee
on $\scri$. Now, since $\scri$ is space-like, the Weyl tensor $C_{abcd}$ is completely determined by its electric and magnetic parts, $E_{ac} := l^{2}\, C_{abcd}\, n^{b}n^{d}$ and $B_{ac} := l^{2}\, {}^{\star}\!C_{abcd}\, n^{b}n^{d}$ and both these fields vanish on $\scri$ because of (\ref{eqcdotniszero}). Therefore, we conclude that the \emph{full} Weyl tensor must vanish on $\scri$: 
\be C_{abcd}\, \hat{=} \, 0. \ee
Note, however, that since this equality holds only on the 3-dimensional surface $\scri$, it does \emph{not} imply that the metric $g_{ab}$ ---or even the metric $q_{ab}$ it induces on $\scri$--- is conformally flat.

Next, we will discuss certain consequences of Bianchi identities that play an important role in the definition of conserved gravitational charges at $\scri$. Let us begin with the contracted Bianchi identity
\begin{equation}
\grad^d C_{abcd} + \grad_{[a} S_{b]c} = 0.
\label{eqcontractedbianchi}
\end{equation}
Using the definition $K_{abcd} = \Omega^{-1}\, C_{abcd}$ of the asymptotic Weyl curvature, it therefore follows that 
\be \lim_{\to\scri}\, \big[\grad_{[a} S_{b]c} + \Omega^{-1}\, C_{abcd} n^{d} \big]\, \hat{=} \,0 .\ee
(\ref{eqsandweyl}) now immediately implies that
\be \label{Tablimit}
\lim_{\to \scri}\, \big[\, \Omega^{-1} \grad_{[a} \left(\Omega\, \tilde{T}_{b]c} \right) - \Omega^{-1} g_{c[a} \tilde{T}_{b]d}\, n^d\,\big] \, \hat{=}\, 0.
\ee
Our boundary conditions on the physical stress energy tensor imply that $\Omega^{-1}\, \hat{T}_{ab}$ has a smooth limit to $\scri$. Therefore, by
writing (\ref{Tablimit}) in terms of quantities that have a limit to $\scri$ we find that
\begin{align}
\lim_{\to \scri} \Big[ 2 n_{[a}\, \Omega^{-1} \hat{T}_{b]c} - g_{c[a}\, \Omega^{-1} \hat{T}_{b]d}\, n^d - \Omega^{-1} \hat{T}_{pq} g^{pq}\, n_{[a} g_{b]c} \Big] \hat{=} 0\, .
\end{align}
The projections of this equation along $n^a n^c q^b_{\;m}$ and $n^a q^b_{\;m} q^c_{\;n}$ will have direct implications to our later discussion:
\begin{align} \label{falloff}
\lim_{\to \scri} \; \Omega^{-1} \hat{T}_{ab} n^a q^b_{\;m} \; &\hat{=} \; 0, \\
\lim_{\to \scri} \; \Omega^{-1} \hat{T}_{ab} q^a_{\;m} q^b_{\;n} \; &\hat{=} \; 0.
\end{align}
Thus, while our basic definition only asked that $\Omega^{-1}\hat{T}_{ab}$ should have a smooth limit to $\scri$, Einstein's equations and Bianchi identities imply that only one component, $\Omega^{-1} \hat{T}_{ab}\, n^{a} n^{b}$, of this limit can be non-zero. 

Finally, recall that the definition of \emph{strongly asymptotically de Sitter space-times} requires that the intrinsic metric $q_{ab}$ on $\scri$ be conformally flat. What is the implication of this additional restriction on properties of the asymptotic Weyl curvature? To analyze this relation, let us first note that because the extrinsic curvature $k_{ab}$ of $\scri$ vanishes in our choice of conformal frame, we can easily express the Riemann tensor $\mathcal{R}_{abcd}$ of the 3-metric $q_{ab}$ in terms of the Riemann tensor $R_{abcd}$ of $g_{ab}$ as follows 
\be \mathcal{R}_{abcd}\, \hat{=} \, q_a{}^{k} q_b{}^{l} q_c{}^{m} q_d{}^{n} R_{klmn}\, .
\ee
Therefore, using (\ref{eqriemannweyl}) and the equation $C_{abcd}\, \hat{=}\, 0$, the Ricci-tensor $\mathcal{R}_{ab}$ of $q_{ab}$ can be expressed in terms of the Schouten tensor $S_{ab}$ as
\be \mathcal{R}_{ab} - \f{1}{4}\mathcal{R} q_{ab}\, \hat{=}\, \f{1}{2}\, q_{a}{}^{m} q_{b}^{n} S_{mn}\, . \ee
Recall that the metric $q_{ab}$ is conformally flat if and only if its (Cotton or) Bach tensor $B_{abc} := D_{[a} (\mathcal{R}_{b]c}-\frac{1}{4} q_{b]c} \mathcal{R})$ vanishes. We can now relate $B_{abc}$ to the asymptotic Weyl curvature as follows: 
\be \label{bach}
B_{abc}\, \hat{=}\, \frac{1}{2} q_a^{\;m} q_b^{\;n} q_c^{\;l} 
D_{[m} S_{n]l}\, \hat{=}\, - \frac{1}{2} q_a^{\;m} q_b^{\;n} q_c^{\;l} K_{mnlp} n^p\, ,
\ee
where, in the last step we have used (\ref{eqcurls}) and denoted the leading order Weyl curvature at $\scri$ by
\be K_{abcd} := \Omega^{-1}\, C_{abcd}\, . \ee
It is easy to check that the right side of (\ref{bach}) vanishes if and only if the leading order magnetic part 
\be \mathcal{B}^{ac}\, :=\,  {}^{\star}\!K^{abcd}\, \mathring{n}_{b} \mathring{n}_{d} =  \f{3}{\Lambda} \,\, {}^{\star}\!K^{abcd}\,  n_{b} n_{d} \ee
vanishes, where $\mathring{n}^{a}$ is the unit future pointing normal to $\scri$. Thus, \emph{the additional restriction that $q_{ab}$ be conformally flat is equivalent to demanding that the asymptotic Weyl curvature $K_{abcd}$ at $\scri$ have no magnetic part.} Now, in electrodynamics, a restriction to Maxwell fields $F_{ab}$ whose magnetic parts $B_{a}$ vanish on a given space-like surface would be severe, as it would cut the space of allowable Maxwell fields by half. We will see in section \ref{s5}, and in \cite{abk2}, that the situation is similar in the gravitational case; the requirement of conformal flatness of $q_{ab}$ severely restricts permissible space-times and this restriction has no physical justification.

\section{Examples}
\label{s3}

This section will discuss several examples of strongly asymptotically de Sitter space-times, both stationary and dynamical, in which the three topologies of $\scri$ discussed in section \ref{s2.1} are realized. Another interesting example ---the Robinson Trautman with a positive cosmological constant--- is discussed in \cite{bicak-padolski}.

\subsection{de Sitter space-time}
\label{s3.1}

Since our Definition 1 is modeled after de Sitter space-time, it provides the simplest example. We will discuss it briefly, mainly to compare and contrast with other examples.

Using the fact that de Sitter space-time can be realized as the unit time-like hyperboloid in 5-dimensional Minkowski space, one can introduce standard `global coordinates' $\tau, \chi, \theta, \phi$ and express the metric as:
\be
\label{eqdSglobal}
{\rm d}\hat{s}^2 = - {\rm d} \tau^2 + l^2\, (\cosh^2 \frac{\tau}{l})\,\, \left( {\rm d} \chi^2 + \sin^2 \chi {\rm d} \omega_2^2 \right)\, ,
\ee
where  ${\rm d}\omega_{2}^{2}$ denotes the unit 2-sphere metric. These coordinates are tailored to a congruence of cosmological observers with spatial sections $\tau = {\rm const}$ that are round 3-spheres. Intuitively, it is clear that future and past infinity $\scri^{\pm}$ should correspond to $\tau=\pm\infty$. The functional form of the radius of the spatial 3-sphere cross-sections ---which diverges in these limits--- leads us to carry out a conformal completion by setting $\Omega=(\cosh (\tau/l))^{-1}$. The rescaled metric then becomes
\be
{\rm d} s^2 = \Omega^2 {\rm d}\hat{s}^2 = - \frac{l^2}{1-\Omega^2} {\rm d} \Omega^2 + l^2 ({\rm d} \chi^2 + \sin^2 \chi {\rm d} \omega_2^2 ).
\ee
Clearly the conformally rescaled metric $g_{ab}$ is well-defined at the boundary $\Omega=0$ which now has two disconnected components $\scri^{\pm}$. Each of these components has topology $\mathbb{S}^{3}$ and is endowed with a metric $q_{ab}$ of a round sphere of radius $l$. Therefore it is geodesically complete. Next, $\nabla_{a} \Omega$ is non-zero at these boundaries because $g^{ab}\,\nabla_{a}\Omega\,\nabla_{b}\Omega \hat{=} - l^{-2}$. Furthermore, since $\hat{g}_{ab}$ is a metric of constant curvature, the stress-energy tensor $\hat{T}_{ab}$ as well as the Weyl tensor $\hat{C}_{abcd}$ vanish identically everywhere in the physical space-time. Therefore, de Sitter space-time satisfies Definition 3 and, this completion makes it \emph{strongly} (and globally) asymptotically de Sitter.

\begin{figure}[]
  \begin{center}
  \vskip-0.4cm
    \includegraphics[width=2.1in,height=2.5in,angle=0]{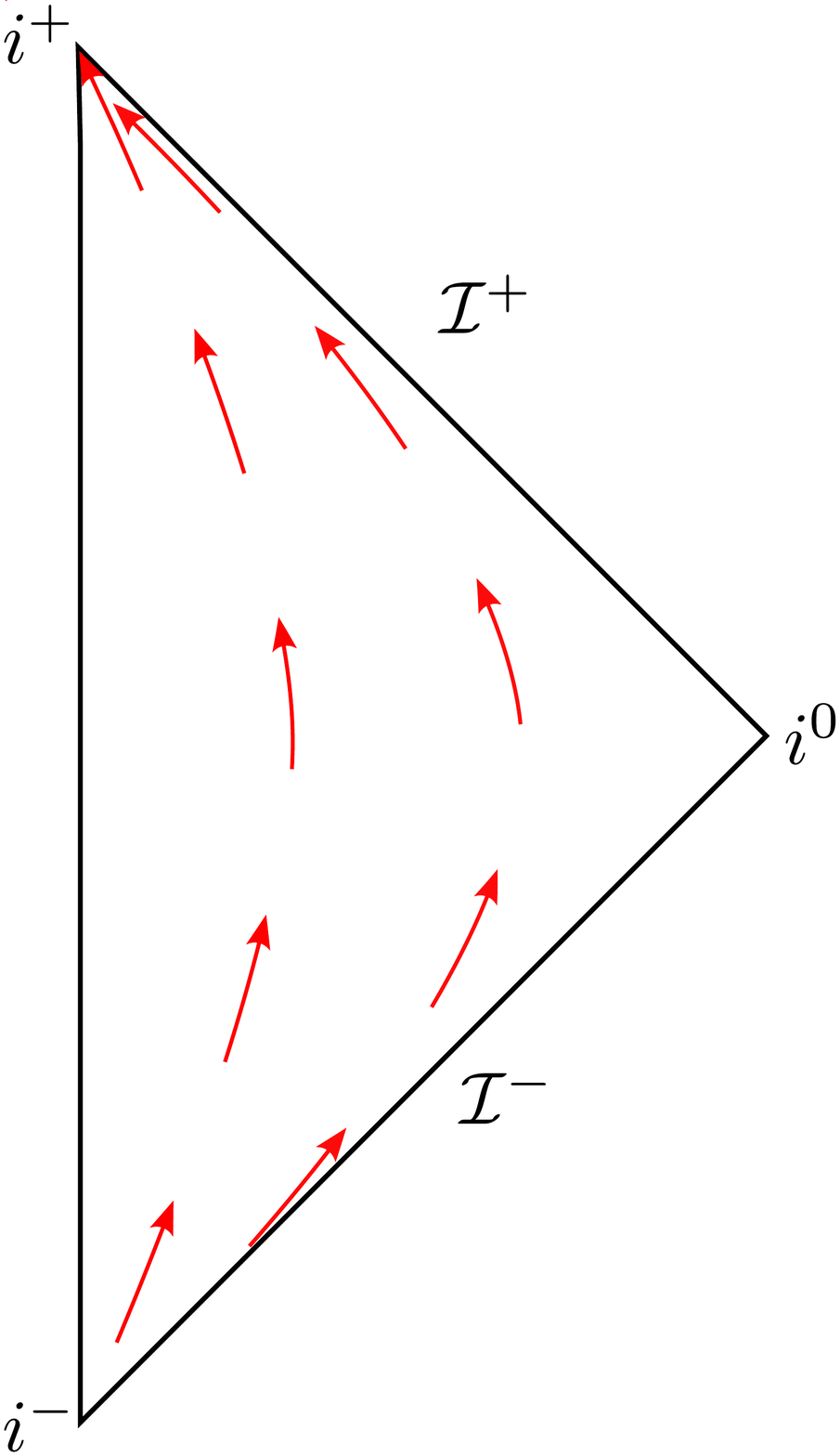}\hskip1.5cm
    \includegraphics[width=2.1in,height=2.5in,angle=0]{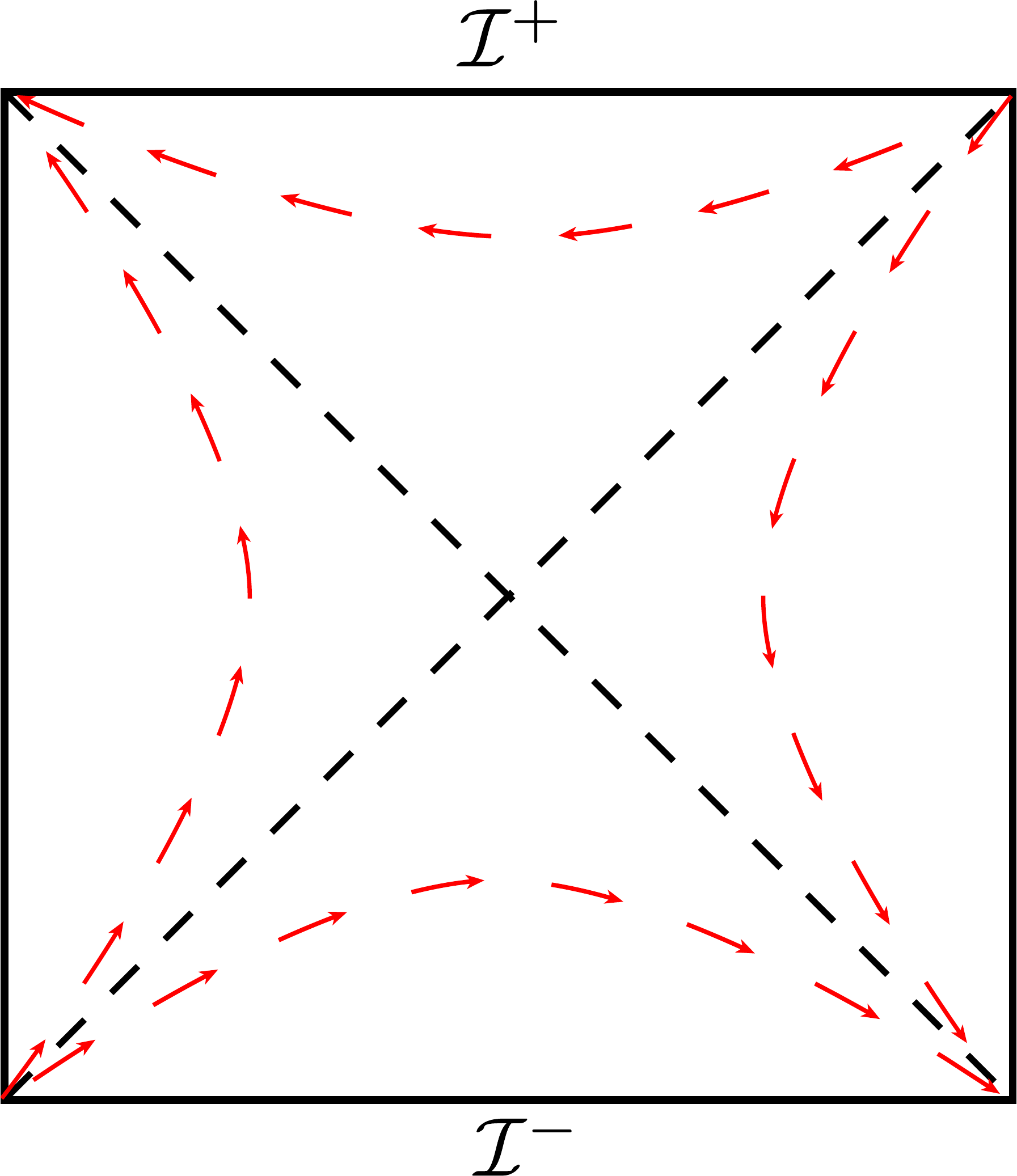}
\caption{Isometries near $\scri$ of Minkowski and de Sitter space-times. \textit{Left Panel:} Minkowski space-time. The time translation Killing fields are time-like in a neighborhood of $\scri$ and null on $\scri$. \textit{Right Panel:} de Sitter space-time. Since $\scri$ is now space-like, \emph{all} Killing fields of de Sitter are space-like near and on $\scri$. The arrows represent a `time translation' which changes its time-like versus space-like character across cosmological horizons.}
\label{ds-mink}
\end{center}
\end{figure}

\emph{Remark:}\\
As in the asymptotically Minkowski case \cite{asymflat}, it is easy to show that every global Killing field of the physical space-time admits an extension to the boundary and is tangential to $\scri$. de Sitter space-time has 10 Killing vector fields. Since $\scri^{\pm}$ are space-like, \emph{every} Killing field is space-like at $\scri^{\pm}$ and in their neighborhoods. This is in striking contrast with the situation in Minkowski space-time (see Fig. \ref{ds-mink}). If matter satisfies the weak-energy condition, then the flux of energy of that field at $\scri^{\pm}$ is necessarily positive in Minkowski space-time because time translations are future pointing null vectors at $\scri^{\pm}$. In de Sitter space-time, by contrast, while one can again single out a 3-parameter family of `time translations', since these Killing fields are space-like on and near $\scri^{\pm}$, the associated flux of de Sitter energy across $\scri^{\pm}$ can have either sign even when the stress-energy tensor satisfies all the standard energy conditions.

\subsection{The Schwarzschild-de Sitter solution}
\label{s3.2}

\begin{figure}[]
  \begin{center}
    \includegraphics[width=3.5in,height=2in,angle=0]{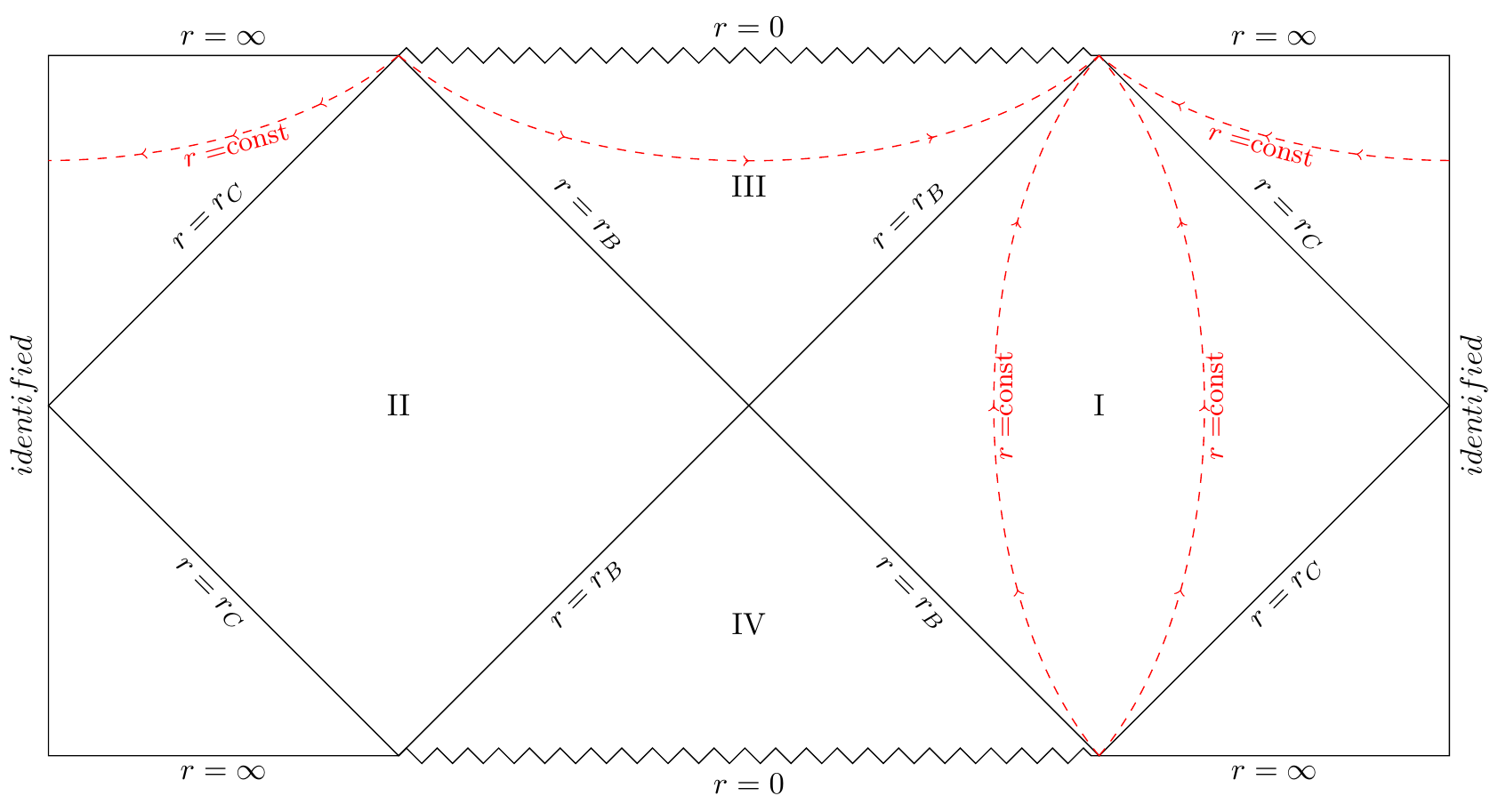}
    
\caption{Conformal diagram of the Schwarzschild-de Sitter space-time. In contrast with the asymptotically flat case, this solution for an eternal black hole admits analytical continuations to the right and left of the diagram, exhibiting an infinite number of black hole and white hole singularities. Therefore, one generally makes an identification. Then the space-time has only one (white hole) singularity in the past and one (black hole) singularity in the future. But now the Cauchy surfaces have a topology $\mathbb{S}^{2}\times \mathbb{S}^{1}$ rather than $\mathbb{S}^{2}\times \mathbb{R}$ as in the asymptotically flat case. Also, we now have additional (cosmological) horizons at $r=r_{c}$ and the `time translation' Killing field (whose orbits are shown in red dashed lines) is space-like near $\scri$.}
\label{sch}
\end{center}
\end{figure}

We will now consider examples that provide prototypes for describing non-dynamical isolated gravitating systems. Let us first consider the Schwarzschild-de Sitter metric in `static coordinates' $t,r,\theta,\phi$ adapted to its four symmetries. While these coordinates do not provide a global chart, they suffice to cover the asymptotic regions. However, in contrast to the $\Lambda=0$ Schwarzschild solution, we now have not only black hole but also cosmological horizons on which the chart breaks down (see Fig. \ref{sch}). Therefore it does not simultaneously cover neighborhoods of both future and past infinity. For definiteness we will focus on the future asymptotic region. The physical metric is given by:
\begin{align}
{\rm d}\hat{s}^2 &= - f(r) {\rm d} t^2 + f(r)^{-1} {\rm d} r^2 + r^2 {\rm d}\omega_2^2,\\
{\rm where}\quad f(r) &= 1 - \frac{2M}{r} - \frac{r^2}{l^2},
\end{align}
and $M$ is the Schwarzschild mass. Set $\Omega = l /r$ (so it is dimensionless) and consider the conformally rescaled metric: 
\be
{\rm d} s^2 : = \Omega^2 {\rm d}\hat{s}^2 = - \left( \Omega^2 - \frac{2M}{l}\Omega^3 - 1 \right) {\rm d} t^2 + \frac{l^2 {\rm d} \Omega^2}{\Omega^2 - \frac{2M}{l} \Omega^3 - 1} + l^2\, {\rm d}\omega_2^2 \, .
\ee
Since the rescaled space-time metric $g_{ab}$ is well-defined at $\Omega=0$ we can extend the physical space-time manifold $\hat{M}$ to a manifold $M$ by attaching the $\Omega=0$ surface which represents $\scri^+$. By inspection, $\scri^{+}$ is space-like and topologically $\mathbb{S}^{2}\times \mathbb{R}$ because it is coordinatized by $r,\theta,\phi$. Since $g^{ab}\,\nabla_{a}\Omega\, \nabla_{b}\Omega \hat{=} - l^{-2}$, clearly $\nabla_a \Omega$ is non-zero at $\scri^{+}$. Next, in this completion, the intrinsic metric $q_{ab}$ on $\scri^{+}$ is given by:
\be 
q_{ab} {\rm d}x^{a} {\rm d}x^{b} = {\rm d}t^{2} + l^{2}\, {\rm d}\omega_{2}^{2}.
\ee 
Since $t \in (-\infty, \infty)$, it is clear that $(\scri^{+},\, q_{ab})$ is geodesically complete. The end $t= -\infty$ represents $i^{o}$ and the end $t=\infty$ represents $i^{+}$. We see explicitly that all the four Killing fields of $\hat{g}_{ab}$ are tangential to $\scri^{+}$, as they must be on general grounds. In particular, the generator of the `time-translation' symmetry is \emph{space-like} on $\scri^{+}$, and indeed in the entire neighborhood of $\scri^{+}$ that our `static' coordinates cover. This is in striking contrast to the $\Lambda=0$ case where $\partial/\partial t$ Killing vector field is null on $\scri$ and time-like in its neighborhood covered by the static chart. Still, in the $\Lambda >0$ case we will show in section \ref{s5.3} that the conserved quantity associated with $\partial/\partial t$ is again the Schwarzschild mass $M$.

Finally, since $\hat{g}_{ab}$ is a solution to source-free Einstein's equations, the condition on the stress energy tensor in Definition 1 is trivially satisfied. Thus we have obtained a conformal completion of $(\hat{M}, \hat{g}_{ab})$ in which it is asymptotically Schwarzschild-de Sitter. Furthermore, $q_{ab}$ can be recast as a conformally flat metric explicitly:
\be q_{ab}{\rm d}x^{a} {\rm d}x^{b} = \frac{l^2}{\tau^2}\,\big({\rm d}\tau^{2} + \tau^{2}\, {\rm d}\omega_{2}^{2}\big) \ee
where $\tau= e^{t/l}$. Hence the space-time is also \emph{strongly} asymptotically de Sitter. \\

\emph{Remark:} Note that if we set $M=0$, the physical metric $\hat{g}_{ab}$ reduces to the de Sitter metric. Therefore, if we again use the conformal factor $\Omega = l/r$, the topology of $\scri^{+}$ would be $\mathbb{S}^{2}\times \mathbb{R}$ and the completion would be conceptually different from that considered in section \ref{s3.1}. This is because the static coordinates underlying this construction ---and hence the conformal completion we obtain by setting $\Omega = l/r$--- are tied to a specific `time translation' Killing field $\partial/\partial t$. In de Sitter space-time, of course, there is no preferred rest frame;\, we have a 10-parameter group of isometries. In particular the ends $i^{+}$ and $i^{o}$ of $\scri^{+}$ of this conformal completion are not left invariant by the full isometry group, whence this completion is unnatural from the perspective of the full structure of the de Sitter space-time.

\subsection{The Kerr-de Sitter solution}
\label{s3.3}

At a conceptual level, the situation with Kerr-de Sitter space-time is completely parallel although now the detailed expressions are significantly more complicated. In Boyer-Lindquist type coordinates the physical metric is given by \cite{carter,matzner}:
\begin{align*}
{\rm d}\hat{s}^2 &= (r^2 + a^2 \cos^2 \theta) \Big[\frac{{\rm d} r^2}{\Delta_r} + \frac{{\rm d} \theta^2}{1+ \frac{a^2}{l^2} \cos^2 \theta} \Big] + \sin^2 \theta \frac{1 + \frac{a^2}{l^2} \cos^2 \theta}{r^2 + a^2 \cos^2 \theta} \Big[\frac{a {\rm d} t - (r^2 +a^2) {\rm d} \phi}{1+ \frac{a^2}{l^2}}\Big]^2 \\
&\qquad - \frac{\Delta_r}{r^2 + a^2 \cos^2 \theta} \Big[\frac{{\rm d} t - a \sin^2 \theta {\rm d} \phi}{1+\frac{a^2}{l^2}}\Big]^2
\end{align*}
where $\Delta_r = -\frac{r^4}{l^2}+ (1 - \frac{a^2}{l^2}) r^2 - 2 M r + a^2$. In the limit $a\to 0$, one recovers the Schwarzschild-de Sitter metric as expected. We can again focus on a future asymptotic region and choose the conformal factor $\Omega=l/r$. In the $(t, \Omega,\theta,\phi)$ coordinates the conformally rescaled metric is then given by:
\begin{align}
{\rm d}s^2 &= \Omega^2 {\rm d}\hat{s}^2\, =\,
l^2 \big(1 + \Omega^2 \frac{a^2}{l^2}\cos^2 \theta \big)\,\, \Big[ \frac{{\rm d} \Omega^2}{-1  + \Big(1-\frac{a^2}{l^2} \Big) \Omega^2 - \frac{2M}{l} \Omega^3 + \frac{a^2}{l^2} \Omega^4} + \frac{{\rm d} \theta^2}{1+\frac{a^2}{l^2} \cos^2 \theta} \Big]\notag \\
&\quad + \sin^2 \theta \frac{1+ \frac{a^2}{l^2}\cos^2\theta}{1+ \Omega^2 \frac{a^2}{l^2}\cos^2 \theta}\,\, \Big[\frac{ \Omega^2 \frac{a}{l}{\rm d} t - l \Big(1+ \Omega^2 \frac{a^2}{l^2}\Big){\rm d} \phi}{1+\frac{a^2}{l^2}}\Big]^2  \notag \\
&\quad - \frac{-1  + \Big(1-\frac{a^2}{l^2} \Big) \Omega^2 - \frac{2M}{l} \Omega^3 + \frac{a^2}{l^2} \Omega^4}{1+ \Omega^2 \frac{a^2}{l^2}\cos^2 \theta}\,\, \Big[\frac{{\rm d} t - a \sin^2 \theta {\rm d} \phi}{1+ \frac{a^2}{l^2}}\Big]^2.
\end{align}
Thus the rescaled metric $g_{ab}$  is well defined at $\Omega=0$ whence we can again use this surface as $\scri^{+}$. Since $g^{ab} \nabla_a \Omega \nabla_b \Omega \hat{=} - l^{-2}$, $\nabla_{a} \Omega$ is nowhere vanishing on $\scri$. The intrinsic 3-metric $q_{ab}$ at $\scri$ is now given by:
\begin{align}
q_{ab}\, {\rm d}x^{a}\, {\rm d}x^{b} &\hat{=} \frac{1}{\left(1 + \frac{a^2}{l^2} \right)^2} {\rm d}t^{2} -  \frac{2 a \sin^2 \theta}{\left(1 + \frac{a^2}{l^2} \right)^2}\, {\rm d}t {\rm d}\phi + \frac{l^2}{1 + \frac{a^2}{l^2} \cos^2 \theta} {\rm d}\theta^{2} + \frac{l^2 \sin^2 \theta}{1+ \frac{a^2}{l^2}} {\rm d}\phi^{2},
\end{align}
and is again conformally flat because its Bach tensor vanishes. Thus the discussion is completely analogous to that in the case of Schwarzschild-de Sitter. In this completion, Kerr-de Sitter is asymptotically Schwarzschild-de Sitter and \emph{strongly} asymptotically de Sitter. We will see in section \ref{s5.3} that the conserved quantity associated with the Killing field $\partial/\partial t$ is  $(1+ (a^{2}/l^{2}))^{-2} M$ and that associated with $\partial/\partial \phi$ is $- (1+ (a^{2}/l^{2}))^{-2} Ma$.

\subsection{The Vaidya-de Sitter solution}
\label{s3.4}

While the examples considered so far are important as they represent physically interesting equilibrium configurations of isolated systems, they do not encode dynamics. The simplest dynamical example is the collapse of a spherical null fluid to form a black hole, or its time reverse, the evaporation of a white hole through emission of a spherical null fluid, described by the Vaidya-de Sitter solutions \cite{vaidya}. While these processes are over-idealized from an astrophysical perspective, the example is conceptually interesting because it offers the first glimpses of the effects of non-trivial dynamics on the structure of cosmological horizons and asymptotic symmetries without recourse to numerical simulations. These lessons will be important for the later part of this program dealing with general isolated systems in full, non-linear general relativity \cite{aads}.

\begin{figure}[]
  \begin{center}
    \includegraphics[width=2.3in,height=1.8in,angle=0]{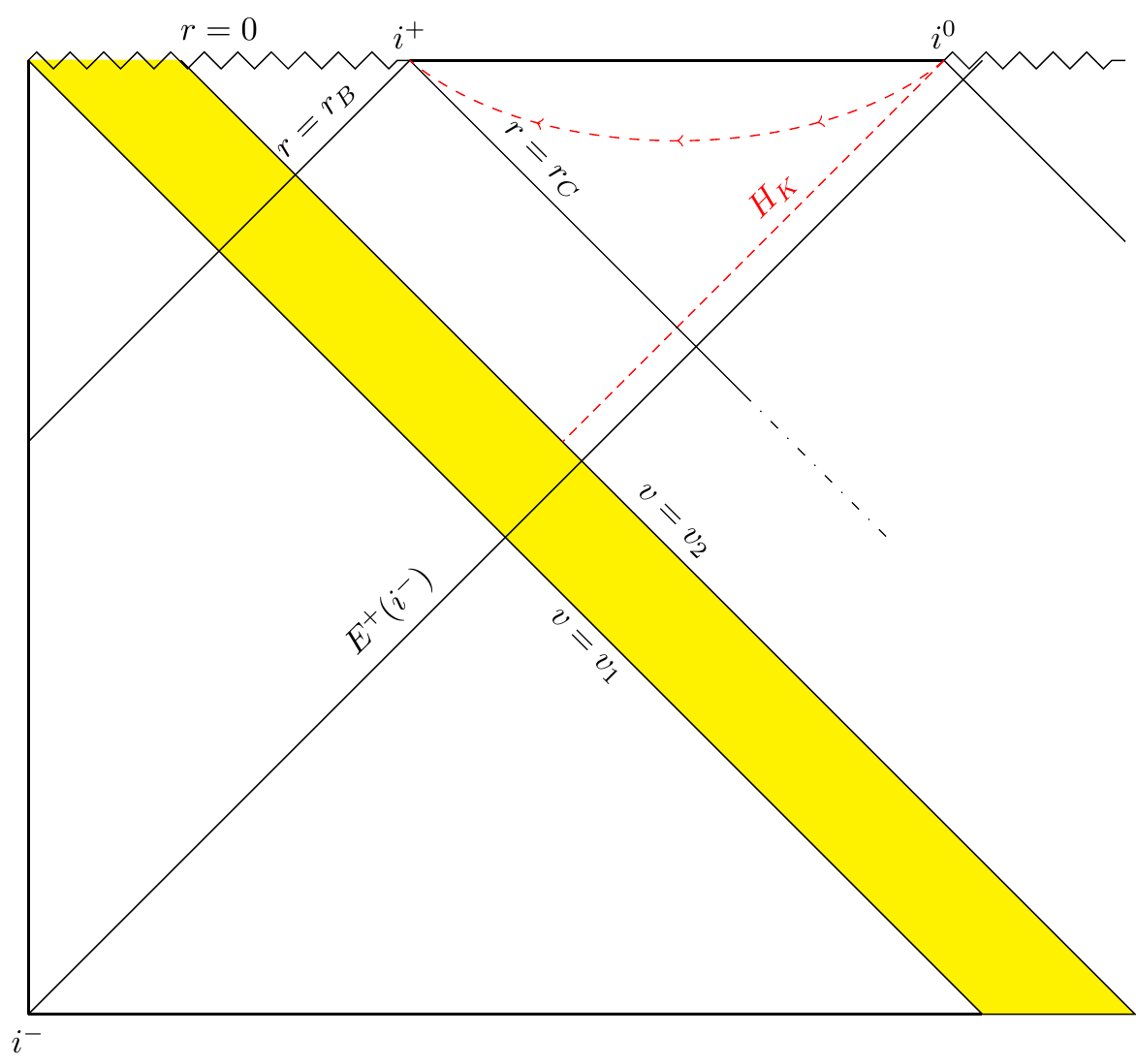}
    
\caption{Conformal diagram of the Vaidya-de Sitter space-time describing the gravitational collapse of an in-falling null fluid from $\scri^{-}$. The in-falling null fluid is indicated by the shaded (yellow) region $v_1 \le v \le v_2$. In contrast to the asymptotically flat case, now the dynamical nature of the space-time geometry modifies the structure even at $\scri^{+}$. While in the Schwarzschild-de Sitter space-time the Killing and the cosmological horizon $E^{+}(i^{-})$ coincide near $\scri^{+}$ they are now distinct; one intersects $\scri^{+}$ and the other meets the singularity. Furthermore, the natural identification that allowed us to consider a single black hole (and the associated white hole) in the Schwarzschild-de Sitter case is no longer available.} 
\label{vaidya}
\end{center}
\end{figure}

Here we will focus on a Vaidya-de Sitter solution that describes black hole formation (see Fig. \ref{vaidya}). In terms of the advanced null coordinate $v$ and the spherical coordinates $r,\theta,\phi$, the metric can be expressed as
\be
 {\rm d}\hat{s}^{2} = - (1 - \frac{2 M(v)}{r} - \frac{r^2}{l^2}) {\rm d}v^{2} + 2 {\rm d}v {\rm d} r + r^2\, {\rm d}\omega_{2}^{2}\, ,
\ee
where the mass function $M(v)$ has the following properties: $M(v) = 0$ for $v<v_1$, it increases monotonically from zero to a value $M$ during the interval $v_1 \leq v \leq v_2$, and $M(v) = M$ for $v>v_2$. $\hat{g}_{ab}$ is a solution to Einstein's equation in presence of a stress-energy tensor 
\be \label{source}
\hat{T}_{ab}  = \frac{\dot{M}}{4\pi r^{2}}\, \nabla_{a}v\, \nabla_{b}v \, ,   \ee
where $\dot{M} = dM/{\rm d} v$. The space-time naturally splits into three regions: de Sitter before collapse ($v< v_1$), dynamical region during collapse ($v_1 \leq v \leq v_2$), and Schwarzschild-de Sitter after collapse ($v_2 < v$). The physical metric $\hat{g}_{ab}$ is spherically symmetric in all three regions. What is the structure of $\scri$?  Since in a neighborhood of $\scri^{+}$ the physical space-time is isometric to the Schwarzschild-de Sitter space-time of mass $M$, the structure of $\scri^{+}$ is the same as that in section \ref{s3.2}. 

On the other hand, because of the incoming radiation, the structure of $\scri^{-}$ is different from that in the Schwarzschild-de Sitter space-time. Let us discuss it in some detail. One can again carry out a conformal completion using $\Omega = l/r$ to attach $\scri^{-}$ as the (past) boundary to the physical space-time. The form (\ref{source}) of the stress-energy tensor implies that not only does $\Omega^{-1}\hat{T}_{ab}$ admit a limit to $\scri^{-}$, but that the limit is in fact zero, in spite of the incoming radiation. Furthermore, although the `time translation' Killing field does not extend to the dynamical region, the affine parameter $t$ of the space-like geodesics orthogonal to the three rotational Killing vectors again runs from $t= -\infty$ to $t=\infty$. Therefore $\scri^{-}$ is also geodesically complete. Next, the asymptotic Weyl curvature $K_{abcd}$ vanishes on the portion of $\scri^{-}$ with $v <v_{1}$, and an explicit calculation shows that it has only a non-zero electric part $\mathcal{E}^{ab}$ for $v >v_{1}$. Thus, in spite of the incoming radiation from $\scri^{-}$, the conformal completion with $\Omega= l/r$ endows the Vaidya space-time with the structure of a strongly asymptotically de Sitter space-time with $\scri^{-}$ of the asymptotically Schwarzschild-de Sitter type. 

The dynamical nature of $\hat{g}_{ab}$ has an interesting consequence for the causal structure of space-time, which will be important to our framework describing general isolated systems in presence of a positive $\Lambda$ \cite{aads}. Let us first examine space-time geometry in the non-dynamical examples discussed in the last two sub-sections. Of particular interest are the event horizons associated with $i^{\mp}$. Their structure is the same in de Sitter and Schwarzschild-de Sitter space-times: the future horizon $E^+(i^-)$ of $i^{-}$ and the past horizon $E^-(i^+)$ of $i^{+}$ are both null 3-surfaces and Killing horizons for the `time translation Killing field', $\partial/\partial t$. (In the de Sitter space-time, this is the `time translation' adapted to the chosen points $i^\mp$ on $\scri^{\mp}$.) These Killing horizons intersect at a bifurcate horizon 2-surface where the `time-translation' Killing vector field vanishes. In our dynamical example, the past event horizon $E^{-}(i^{+})$ of $i^{+}$ is again a Killing horizon because it lies entirely in the $v > v_{2}$ region. But since $E^{+}(i^{-})$ does intersect the dynamical region, it is no longer a Killing horizon to the future of the $v=v_{1}$ surface. In the region $v >v_{2}$ we do have a Killing horizon $H_{K}$ for the `time-translation' of the Schwarzschild-de Sitter metric. Like $E^{+}(i^{-})$, it is a null surface. \emph{However, $H_{K}$ lies strictly to the future of $E^{+}(i^{-})$} (see Fig. \ref{vaidya}). The Killing field is now \emph{transversal} to the portion of $E^{+}(i^{-})$ that lies to the future of $v=v_{2}$. This split between $E^{+}(i^{-})$ and the Killing horizon $H_{K}$ has interesting implications for the conceptual framework describing non-stationary isolated systems with positive $\Lambda$ both in the classical \cite{aads} and quantum regimes.

\subsection{Friedmann-Lema\^itre cosmology}
\label{s3.5}

The notion of asymptotically de Sitter space-times is useful not only to the study of isolated systems but also in the description of the late time behavior of the universe in Friedmann-Lema\^itre cosmology with positive $\Lambda$. In this case, Einstein's equations inform us that if matter obeys the strong energy condition, then the result of the expansion is that the cosmological constant dominates at late times, irrespective of how small its value is. Current observations imply that today the source of the Hubble parameter has two predominant components, modeled by dust ($\sim 30\%$) and $\Lambda$ ($\sim 70\%$). Since the expansion of the universe dilutes the energy density of dust as $a^{-3}$ (where $a$ is the co-moving scale factor) and the energy density of $\Lambda$ remains constant, given sufficient time, the universe will be naturally driven to become an asymptotically de Sitter space-time. We will now make these qualitative considerations more precise.

Let us first use Einstein's equations together with observational inputs to construct a space-time metric to describe the late stages of the large scale dynamics of our universe. For spatially flat, $k=0$ universe, the physical metric has the form
\begin{equation}
{\rm d} \hat{s}^2 = - {\rm d} t^2 + a^2(t) \left({\rm d} x^2 + {\rm d} y^2 + {\rm d} z^2 \right)
\end{equation}
and Einstein's equations allow us to relate the matter content of the universe with the time dependence of the scale factor $a$:
\begin{align} \label{eqhubbel}
H_0 (t - t_\star) & = \int_{a_\star}^{a(t)} \! \frac{{\rm d} \tilde{a}}{\sqrt{\frac{\Omega_{r,0} a_0^4}{\tilde{a}^2}+ \frac{\Omega_{d,0} a_0^3}{\tilde{a}}+\Omega_{\Lambda,0} \tilde{a}^2 + (1 - \Omega_c)}}\, 
\end{align}
Here $H = \dot{a}/a$ is the Hubble parameter; the subscript $0$ refers to today's values and the subscript $\star$ to values at any chosen `initial' time; and $\Omega_i = (8\pi G/3H^{2})\, \rho_{i}$ are the fractional density parameters, with the subscript $r$ referring to radiation, $d$ to dust, and $c$ to the critical density $\rho_c := ({3H^2}/{8 \pi G})$. Taking input from current observations for the values of the density parameters: $\Omega_{r,0} \sim 0$, $\Omega_{d,0} \sim 0.3$, $\Omega_{\Lambda,0} \sim 0.7$ and $\Omega_c \sim 1$, and setting $t_{\star}\gg t_{0}$, we obtain%
\be
\label{eqhubblesoln}
H_0 (t - t_\star) = \frac{2}{3 \sqrt{0.7}} \ln\big[\frac{\sqrt{0.7 a^3} + \sqrt{0.3 a_0^3 + 0.7 a^3}}{\sqrt{0.7 a_{\star}^3} + \sqrt{0.3 a_0^3 + 0.7 a_{\star}^3}}\big]. 
\ee
Since the expansion is dominated by $\Lambda$ at late times, the leading-order behavior of the scale factor is expected to be exponential. This is borne out by inverting (\ref{eqhubblesoln}) to obtain the scale factor as a function of time as follows:
\be a(t) = \frac{e^{-\beta \Delta t}}{(2.8)^{1/3}} [0.3 a_0^3 (1 - e^{3\beta \Delta t})^2 + 1.4 a_{\star}^3 (1 + e^{6\beta \Delta t}) - 2\sqrt{(0.7 a_{\star}^3)(0.3 a_0^3 + 0.7 a_{\star}^3)}(1-e^{6\beta \Delta t})]^{1/3}
\ee
where the parameter $\beta: = \sqrt{0.7} H_0$ conveniently captures the residual effect of the presence of dust, and $\Delta t := (t - t_{\star})$. At late times, the scale factor simplifies to:
\be a(t) \to \Gamma a_{\star} e^{\beta \Delta t} 
\ee
where $\Gamma = \left[ \frac{1}{2} + \frac{0.3 a_0^3}{4* 0.7 a_{\star}^3} + \frac{1}{2}\sqrt{1 + \frac{0.3 a_0^3}{0.7 a_{\star}^3}}\right]^{1/3}$. Note that in the absence of dust we would have de Sitter space-time where $\Gamma = 1$ and $\beta = H_0$.

For the conformal completion we wish to carry out, it is convenient to re-express the scale factor as a function of the conformal time $\eta$, related to the co-moving time $t$ via  ${\rm d} \eta = {\rm d} t / a$: 
\be
a(\eta) = -\frac{1}{\beta \eta}.
\ee

Thus, at late times, we can write the physical metric as 
\begin{align}
{\rm d} \hat{s}^2 &= \frac{1}{\beta^2 \eta^2} \Big( - {\rm d} \eta^2 + {\rm d} x^2 + {\rm d} y^2 + {\rm d} z^2 \Big). \label{metric}
\end{align}
We can now carry out the conformal completion to verify if conditions in Definition 2 are met. The form (\ref{metric}) suggests that we set $\Omega = -\beta\, \eta$ so that the conformally rescaled metric is given by:
\be {\rm d}s^{2}\, =\, - \beta^{-2} {\rm d} \Omega^2 + {\rm d} x^2 + {\rm d} y^2 + {\rm d} z^2. \label{metric}
\ee
On $\scri$, where $\Omega = 0$, $\nabla_{a}\Omega$ is non-zero because $g^{ab}\nabla_{a}\Omega\nabla_{b}\Omega\, \hat{=} - \, \beta^{2}$. The physical metric $\hat{g}_{ab}$ satisfies Einstein's equations with $\hat{T}_{ab} = \hat{\rho} \hat{u}_a \hat{u}_b$ with $\hat{u}^a$ the unit 4-velocity of a co-moving observer and $\hat{\rho}= \hat{\rho}_d = 0.3 \, \hat{\rho}_{c,0}\,(a_{0}^{3}/{a(t)^3})$. Therefore, in terms of fields which have well-defined limits to $\scri^{+}$, we have  $\Omega^{-1}\, \hat{T}_{ab} = {\rm const}\,\, u_{a} u_{b}$ (with $g^{ab}u_{a}u_{b} =-1$), which admits a smooth limit to $\scri^{+}$. Finally, by inspection, the induced metric on $\scri^{+}$ is:
\be q_{ab}{\rm d}x^{a}{\rm d}x^{b} = {\rm d}x^{2} + {\rm d}y^{2} +{\rm d}z^{2}. \ee
Hence $\scri^{+}$ is geodesically complete and (conformally) flat. The fact that it is coordinatized by $x,y,z$ shows that its topology is $\mathbb{R}^{3}$. Thus our conformal completion makes $(\hat{M}, \hat{g}_{ab})$ \emph{strongly asymptotically de Sitter in a Poincar\'e patch.} This is just what one would expect because, since the future event horizon $E^{+}(i^{-})$ of $i^{-}$ in de Sitter space-time corresponds to the big bang singularity, $(\hat{M}, \hat{g}_{ab})$ is conformally isometric to the expanding Poincar\'e patch, the `upper triangle', of de Sitter space-time. Finally, if we set $\hat{\rho} =0$, locally the solution reduces to the de Sitter space-time. Therefore we could have used $\Omega= -\beta \, \eta$ also in that case. The resulting conformal completion would be inequivalent to the natural conformal completion discussed in section \ref{s3.1} where $\scri^{+}$ is topologically $\mathbb{S}^{3}$ and includes the point $i^{o}$ at spatial infinity. 

\begin{figure}[]
  \begin{center}
    \includegraphics[width=3.5in,height=2.8in,angle=0]{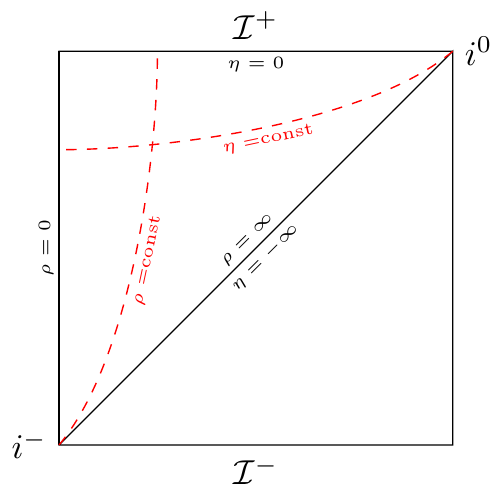}
    
\caption{Conformal diagram of the Friedmann-Lema\^itre space-time with positive $\Lambda$. This space-time corresponds only to the Poincar\'e patch of de Sitter space-time because of the big-bang singularity along the event horizon $E^{+}(i^{-})$ (i.e., $\eta = -\infty$, where $\eta$ is the conformal time).} 
\label{poincare}
\end{center}
\end{figure}

We conclude by noting that the `genuinely dynamical' nature of this space-time distinguishes it from de Sitter space-time in three respects. First, as noted above, even though the conformally rescaled metric $g_{ab}$ at $\scri^{+}$ is the same as that in de Sitter space-time, the conformal factor $\Omega = -\beta \, \eta = -\sqrt{0.7} \, H_0 \, \eta$ retains a memory of the matter content. Second difference is more important: whereas the area of the event horizon $E^{-}(p)$ of any point $p$ on $\scri^{+}$ is constant in de Sitter space-time, it increases monotonically as one approaches $\scri^{+}$ in the Friedmann-Lema\^itre space-time because of the matter content. Finally, this space-time has a big-bang singularity at $t=0$ or $\eta = - \infty$. Therefore it corresponds only to the `upper triangle' of de Sitter space-time (see Fig. \ref{poincare}).\\

\emph{Remarks:}\\
1) Recall from section \ref{s2.2} that although in Definition 1 the explicit requirement is only that $\Omega^{-1} \hat{T}_{ab}$ should have a smooth limit to $\scri$, field equations and Bianchi identities ensure that the space-space and space-time components of this limit vanish and only the time-time component can be non-zero. If the matter consists of Yang-Mills fields or a null fluid of the Vaidya-de Sitter solution, or radiation filled Friedmann-Lema\^itre cosmology, even this component vanishes. In the definition of asymptotically Minkowski space-times one routinely requires that $\Omega^{-2} \hat{T}_{ab}$ have a smooth limit to $\scri$. We did not impose this stronger fall-off condition because in the dust filled Friedmann-Lema\^itre solution we just discussed, the limit of the time-time component of $\Omega^{-1} \hat{T}_{ab}$ is smooth but non-zero. 

2) We focused on the Friedmann-Lema\^itre model because it is used very widely in the contemporary cosmological literature. However, we expect that the early investigations by Wald \cite{waldcosmology} of Bianchi models will provide additional examples once appropriate restrictions are made on matter fields. We also expect that the much more general results obtained recently by Ringstr\"om \cite{hr} on absence of cosmological hair will provide a large class of examples satisfying our Definition 2 of asymptotically de Sitter space-times. However, a detailed analysis is necessary to relate the results obtained in these references in physical space-times to extract the precise asymptotic behavior of various fields after appropriate conformal completions.

\section{Asymptotic symmetries}
\label{s4}

Given a set of boundary conditions, the asymptotic symmetry group $\G = {\rm Diff}_{\infty}(M)/{\rm Diff}_{\infty}^{0}(M)$ is the quotient of the group ${\rm Diff}_{\infty}(M)$ of diffeomorphisms on the physical space-time $(\hat{M}, \hat{g}_{ab})$ that preserve the boundary conditions by its sub-group ${\rm Diff}_{\infty}^{0}(M)$ of diffeomorphisms that are asymptotically identity. At the infinitesimal level, elements of the Lie algebra $\g$ of $\G$ can be naturally represented by vector fields $\xi^{a}$ on $\scri$,\, motions along which preserve the universal structure ---the structure that is shared by \emph{all} space-times satisfying the given boundary conditions. As we noted in section \ref{s1}, the situation with asymptotic symmetries is rather subtle for $\Lambda >0$. Therefore we will first briefly recall the asymptotic symmetries for $\Lambda=0$, and then discuss the $\Lambda >0$ case. This detour will also serve to bring out the reason behind the surprising difference in the asymptotic symmetry groups in the two cases.

\subsection{Asymptotically Minkowski space-times}
\label{s4.1}

The definition of asymptotically Minkowski space-times is the same as Definition 2, but now $\Lambda$ is set to zero, $\Omega^{-2} \hat{T}_{ab}$ is required to have a smooth limit, and the completeness requirement is adapted to the null character of $\scri$ \cite{asymflat}. The resulting structure at $\scri$ can be summarized as follows. Since $\scri$ is null, the null normal $n^{a}$ is now also tangential to $\scri$ and the intrinsic metric $q_{ab}$ on $\scri$ is degenerate, with signature 0,+,+.\, Under a conformal rescaling $\Omega \to \Omega^{\prime} = \omega \Omega$, we have $n^{\prime\,a} \, \hat{=} \,  \omega^{-1}n^{a}$ and $q^{\prime}_{ab} \, \hat{=} \,  \omega^{2} q_{ab}$. Using this freedom, again one can always pass to a conformal frame in which $\nabla_{a} n^{a} \, \hat{=} \,  0$.  The conformal freedom is then reduced to $\Omega^{\prime} \, \hat{=} \,  \omega \Omega$ where $\mathcal{L}_{n}\, \omega \, \hat{=} \, 0$. The completeness restriction on $\scri$ is the requirement that the vector field $n^{a}$ be complete in any of these `divergence-free' conformal frames. Field equations imply that in these conformal frames, the pull-back to $\scri$ of $\nabla_{a}n_{b}$ vanishes, whence $\Lie_{n}\, q_{ab} \, \hat{=} \, 0$. The integral curves of $n^{a}$ are referred to as \emph{generators of} $\scri$. Finally, one can pass to the space $S$ of generators of $\scri$. $S$ is required to have topology $\mathbb{S}^{2}$ to capture the fact that one can move away from the isolated system along null rays in any angular direction. Therefore the topology of $\scri$ is $\mathbb{S}^{2}\times \mathbb{R}$ and the affine parameter $u$ of $n^{a}$ spans the full interval $(-\infty, \, \infty)$ in the $\mathbb{R}$-direction. 

The universal structure is therefore given by pairs $(q_{ab}, n^{a})$ of fields on a 3-manifold $\scri$ with topology $\mathbb{S}^{2}\times \mathbb{R}$ such that: i) $n^{a}$  is complete; ii) $q_{ab}$ is a degenerate metric of signature $0,+,+$ with $q_{ab}n^{b} \, \hat{=} \, 0$ and $\mathcal{L}_{n}\,q_{ab} \, \hat{=} \, 0$; and, iii) any two pairs $(q_{ab}, n^{a})$ and $(q^{\prime}_{ab}, n^{\prime}{}^{a})$ are related by  $q^{\prime}_{ab} \, \hat{=} \,  \omega^{2} q_{ab}$ and $n^\prime{}^{a} \, \hat{=} \,  \omega^{-1}n^{a}$ for some $\omega$ satisfying $\Lie_{n}\omega \, \hat{=} \,  0$. Because 2-spheres carry a unique conformal structure, the metrics $q_{ab}$ in this collection are all conformal to a unit 2-sphere metric.

The BMS group $\B$ is the group of diffeomorphisms preserving this universal structure. $\B$ is substantially smaller than $\Diff$ and furthermore has rich, physically interesting structure because of two reasons. First, 
the BMS symmetries must \emph{preserve the ruling of $\scri$ by its null normals}. Therefore, a BMS vector field $\xi^{a}$ must satisfy $\Lie_{\xi} n^{a} \, \hat{=} \,  \alpha n^{a}$ for some function $\alpha$ (satisfying $\Lie_{n} \alpha \, \hat{=} \, 0$). This implies that vector fields $\xi^{a} = fn^{a}$ (with $\Lie_{n} f\, \hat{=} \, 0$) form a Lie ideal of the BMS Lie algebra. This is the infinite-dimensional sub Lie algebra $\LS$ of \emph{BMS supertranslations.} Thus, because the diffeomorphisms generated by the BMS vector fields must preserve the natural ruling of $\scri$, $\B$ is smaller than $\Diff$ and has a semi-direct product structure. Next, note that the condition $\Lie_{\xi} n^{a} \, \hat{=} \,  \alpha n^{a}$ also implies that \emph{every} BMS vector field $\xi^{a}$ can be projected to a vector field $\bar{\xi}^{a}$ on the 2-sphere $S$ of generators of $\scri$ which then characterizes the element of the quotient $\LBMS / \LS$ it naturally defines. Furthermore, the condition that the pairs $(q_{ab}, n^{a})$ be preserved by the BMS action implies that $\bar{\xi}^{a}$ is a conformal Killing field on the space $S$ of generators of $\scri$ (equipped with metrics $q_{ab}$). The second key factor that dictates the structure of the BMS group is that \emph{the 2-sphere has a unique conformal structure}. Therefore, the quotient  $\LBMS/\LS$ is just the Lie algebra of conformal isometries of a round 2-sphere, {which turns out to be isomorphic to the Lorentz Lie algebra in 4 dimensions}. This is why $\B$ is the semi-direct product, $\B = \S\, \ltimes \L$, of the group $\S$ of supertranslations with the Lorentz group $\L$. Finally, the unique conformal structure of $\mathbb{S}^{2}$ also leads to the non-trivial result that $\B$ admits a unique 4-dimensional normal sub-group of translations $\T$ \cite{sachs}.
 
\subsection{Asymptotically de Sitter space-times}
\label{s4.2}

Let us now consider the $\Lambda >0$ case. Although the definition of asymptotically Minkowski space-times is completely parallel to that of asymptotically de Sitter space-times, the fact that $\Lambda$ is now positive rather than zero has a drastic effect on the structure of the asymptotic symmetry group. First, as we saw in section \ref{s2.2}, now $\scri$ is space-like. Therefore $n^{a}$ is no longer tangential to $\scri$, whence $\scri$ no longer carries a natural ruling. Consequently the presence of $n^{a}$ does not restrict the diffeomorphisms on $\scri$ in any way. Indeed, $\Diff$ does not even have a well-defined action on $n^{a}$ now. 

In any admissible conformal completion of the physical space-time, $\scri$ does carry an intrinsic positive definite metric $q_{ab}$ and we have the rescaling freedom $q_{ab} \to q^{\prime}_{ab} = \omega^{2} q_{ab}$. The second major difference from asymptotically Minkowski $\scri$ is that (in contrast to the 2-sphere that featured in that analysis) Riemannian \emph{3}-manifolds do not carry a fixed conformal structure. For definiteness, let us fix the topology of $\scri$ to be $\mathbb{S}^{3}$, although our arguments extend to other topologies as well. Conditions in the definition do not restrict $q_{ab}$ to be conformally related to the unit 3-sphere metric $\qo_{ab}$. In particular, while $\qo_{ab}$ has vanishing Bach tensor (and is therefore conformally flat), there is no a priori reason for the metric $q_{ab}$ obtained in a conformal completion of a general, globally asymptotically de Sitter space-time to share this property. Indeed, Friedrich's global analysis of stability of de Sitter space-time \cite{hf} shows that there are globally asymptotically de Sitter space-times in which the metric $q_{ab}$ on $\scri$ can lie anywhere in an open ball around $\qo_{ab}$ in a certain function space. (This point will be made explicit using gravitational perturbations of de Sitter space-time in \cite{abk2}.) Thus a priori the universal structure at $\scri$ would consist of all 3-metrics $q_{ab}$ on $\mathbb{S}^{3}$ of signature +,+,+. Since the asymptotic symmetry group $\G$ consists of all diffeomorphisms that preserve the universal structure, the group would be all of $\Diff$. This group does not admit any preferred `translation' and `rotation' sub-groups that can be used to define energy-momentum and angular momentum in a canonical fashion. Thus, the rich structure made available by the BMS group $\B$ in the asymptotically Minkowski space-time to define Bondi charges and fluxes \emph{disappears} once there is a positive $\Lambda$, \emph{however small.}

\emph{Remarks:}\\  \indent
1) Friedrich's results \cite{hf} only show that $q_{ab}$ can be any metric in an \emph{open neighborhood} of $\qo_{ab}$, rather than any metric on $\mathbb{S}^{3}$. However, his goal was to establish \emph{global} stability of de Sitter space-time while here the goal is much more modest; now only the asymptotic behavior near $\scri$ is relevant. Furthermore, while Friedrich restricted himself only to Maxwell and Yang-Mills fields (to control the estimates needed for his global result), in this asymptotic analysis we allow general matter fields, subject only to the requirement that $\Omega^{{-1}}\hat{T}_{ab}$ has a well-defined limit to $\scri$. Because these requirements are significantly weaker, it is reasonable to expect that any metric $q_{ab}$ on $\mathbb{S}^{3}$ will be realized as the intrinsic metric on $\scri$ of some asymptotically de Sitter space-time. 

2) What would happen if we have $\Lambda <0$? Now $\scri$ would be time-like. But except for the difference in the signature of $q_{ab}$ all our considerations will still apply and the asymptotic symmetry group would again be $\Diff$ for space-times satisfying only the conditions of Definition 2 but with $\Lambda <0$. However, as explained in section \ref{s1}, because $\scri$ is now time-like, to obtain well-defined evolution, we are naturally led to impose suitable boundary conditions on fields at $\scri$ in addition to the conditions of Definition 2. These conditions significantly reduce the symmetry group (see, e.g., \cite{swh,aaam,aasd}).

\subsection{Strongly asymptotically de Sitter space-times} 
\label{s4.3}

Let us now consider only those space-times that satisfy Definition 3, i.e., require that the metric $q_{ab}$ be conformally flat, or equivalently, the magnetic part $\mathcal{B}^{ab}$ of the leading order Weyl curvature be zero on $\scri$. We will discuss the three topologies introduced in section \ref{s2.1} in turn.

\subsubsection{$\mathbb{S}^{3}$ topology}
\label{s4.3.1}

Now the universal structure consists of the 3-manifold $\scri$ with $\mathbb{S}^{3}$ topology, equipped with the class of conformally flat metrics $q_{ab}$ of signature +,+,+. Therefore now the group $\G$ of asymptotic symmetries is just the group of conformal isometries of any one metric $q_{ab}$ in this class. It is well-known that this group is isomorphic with the de Sitter group $\Gds \equiv {\rm SO(1,4)}$. Thus, conformal flatness of $q_{ab}$ is a strong requirement as it reduces the infinite dimensional $\Diff$ to a 10-dimensional group. One can give a convenient description of $\G$ using the unit round metric $\qo_{ab}$ in this conformal class:
\begin{align}
\label{canonicals3}
 \qo_{ab}\, {\rm d} x^a {\rm d} x^b = {\rm d} \chi^2 + \sin^2 \chi \,\,( {\rm d} \theta^2 + \sin^2 \theta {\rm d} \phi^2 ),
\end{align}
where $\chi \in \left[0,\pi \right]$, $\theta \in \left[0,\pi \right]$ and $\phi \in \left[0, 2 \pi \right)$.  The Lie algebra $\g$ of $\G$ is spanned by conformal Killing fields on $(\scri, \qo_{ab})$. The 6 Killing vectors of $\qo_{ab}$ provide us with the Lie algebra ${\rm so(4)}$ of ${\rm SO(4)}$ which naturally splits into two ${\rm so(3)}$ sub-algebras: ${\rm so(4)} = {\rm so(3)}_{\rm L} \oplus {\rm so(3)}_{\rm R}$. One can choose a basis in the 10-dimensional Lie algebra ${\rm so(1,4)}$ such that the remaining 4-dimensional space is spanned by \emph{`pure'} conformal Killing fields $C^{a}$ on $(\scri, \qo_{ab})$, i.e., vector fields on $\scri$ which satisfy not only $\Lie_{C}\, \qo_{ab} \, \hat{=} \,  2\alpha\, \qo_{ab}$ for some smooth function $\alpha$ but also $\mathring{D}_{[a} C_{b]} \, \hat{=} \,  0$. (Recall that the Killing fields $K^{a}$ satisfy $\Do_{(a} K_{b)} =0$; hence the terminology `pure' conformal for the vector fields $C^{a}$.) 

Let us embed $(\scri, \qo_{ab})$ as the unit 3-sphere in (an abstractly defined) 4-dimensional Euclidean space $(\mathbb{R}^{4}, \eo_{IJ})$. Then, it turns out that there is a natural 1-1 correspondence between the 10 Killing fields of $(\mathbb{R}^{4}, \eo_{IJ})$ and elements of $\g$. Each of the 6 Killing fields of $\qo_{ab}$ is of course just the restriction of a rotational Killing field of $(\mathbb{R}^{4}, \eo_{IJ})$ to $\scri$, and is therefore labelled by a 2-form $\Ko_{IJ}$ on $\mathbb{R}^{4}$:
\be K^{a} = \qo^{aI}\, \Ko_{IJ}\, x^{J}\, , \ee
where $\qo^{aI}$ is the projection operator on the unit 3-sphere and $x^{J}$ are the position vectors of points in $\mathbb{R}^{4}$. The pure conformal Killing fields on $\scri$ turn out to be just the projections of the 4 translational Killing fields $\Co_{I}$ on $(\mathbb{R}^{4}, \eo_{IJ})$ to $\scri$:
\be C^{a} =  \qo^{a}{}_{I}\, \Co^{I}. \ee

Thus, each form $\Ko_{IJ}$  on $\mathbb{R}^{4}$ defines a Killing field $K^{a}$ and each vector $\Co^{I}$ defines a `pure' conformal Killing field $C^{a}$ on $(\scri, \qo_{ab})$. The commutation relations are then given by:
\begin{align}\label{brackets} [K,\, K^{\prime}] &= K^{\prime\prime}\, ,\quad  {\rm where} \quad \Ko^{\prime\prime}_{I}{}^{J} = \Ko_{I}{}^{L} \Ko^{\prime}_{L}{}^{J} - \Ko^{\prime}_{I}{}^{L} \Ko_{L}{}^{J}\nonumber \\
[K,\, C] &= C^{\prime}\, , \quad  {\rm where} \quad  \Co^{\prime}_{I} = - \Ko_{IJ}\,  \Co^{J}\,\, , {\rm and} \nonumber\\ 
[C,\, C^{\prime}] &= K\, ,   \quad  {\rm where} \quad \Ko_{IJ} = 2\,\Co^{\prime}_{[I} \Co_{J]} \, . \end{align}
%
%

This is a convenient basis for calculations. For example, the 4 `pure' conformal Killing fields $C^{a}$ are generally taken to represent `translations' in $\g$, and are used to define the de Sitter energy-momentum, analogs of the more familiar energy-momentum 4-vectors in asymptotically Minkowski space-times. However, note that while the first two brackets in (\ref{brackets}) mimic the familiar commutation relations between rotations and between rotations and translations, the last bracket does not: while the 4 translations $\Co^{I}$ commute on $(\mathbb{R}^{4}, \eo_{IJ})$, the 4 `pure' conformal Killing fields $C^{a}$ in $\g$ do not.

\subsubsection{$\mathbb{R}^{3}$ topology}
\label{s4.3.2}

As we just saw, when the topology of $\scri$ is $\mathbb{S}^{3}$, the full de Sitter group $\Gds$ constitutes the asymptotic symmetry group $\G$. For other topologies, the local structure is the same. In particular, when $\mathcal{B}^{ab} \, \hat{=} \, 0$, the intrinsic metric on $\scri$ is again conformally flat and, given any conformal completion, we are led to consider the 10 conformal Killing fields of $\qo_{ab}$.  Recall however that in Definition 2 of asymptotically de Sitter space-times we also required that $(\scri, q_{ab})$ be complete. Therefore the question is if the 3-manifold $\scri$ we began with continues to be complete also with respect to the image of $q_{ab}$ under diffeomorphisms generated by all 10 conformal Killing fields. When $\scri$ is topologically $\mathbb{S}^{3}$ as in section \ref{s4.3.1}, this is assured by compactness of $\scri$. But in non-compact topologies this issue has to be analyzed case by case. The asymptotic symmetry group can be smaller if $\scri$ fails to remain complete with respect to the image of $q_{ab}$ under some conformal isometries. We will find that this does happen. When the topology is $\mathbb{R}^{3}$, the group is reduced to a 7-dimensional sub-group and when it is $\mathbb{S}^{2} \times \mathbb{R}$ to a 4-dimensional sub-group.

Before entering the calculations of completeness, it is instructive to return to the example of de Sitter space-time with a conformal completion that bestows $\mathbb{R}^3$ topology on $\scri$ (as in section \ref{s3.5}). In this case, one focuses only on the Poincar\'e patch of de Sitter space-time,  i.e., on  the causal future of a chosen point $i^{-}$ on $\scri^{-}$, which represents the past time-like infinity of a family of observers in de Sitter space-time (see Fig. \ref{poincare}). This patch is covered by coordinates $\eta, x,y,z$, introduced in section \ref{s3.5}, with $x,y,z$ assuming their full range on $\mathbb{R}^{3}$ and the conformal time $\eta \in (-\infty,0)$. The past boundary of this region is the event horizon $E^{+}(i^{-})$ of $i^{-}$ which is \emph{not} part of the Poincar\'e patch because $\eta = -\infty$ there. Now, since the metric in the Poincar\'e patch is just the de Sitter metric, \emph{locally} it admits 10 Killing fields. However, since our space-time is only a part of the de Sitter space-time, only those isometries are permissible that map the Poincar\'e patch to itself. In terms of Killing fields, then, only those de Sitter Killing fields are now permissible that are tangential to $E^{+}(i^{-})$. Geometrically, it is simplest to characterize this restriction by embedding de Sitter space-time as a hyperboloid $\mathcal{H}$ in a 5-dimensional Minkowski space $\mathcal{M}_{5}$. The 10 Killing fields of de Sitter space-time are just the Lorentz generators in $\mathcal{M}_{5}$ and the event horizon $E^{+}(i^{-})$ is realized as the intersection of $\mathcal{H}$ with a 4-dimensional null hyperplane $\mathcal{N}$ passing through the origin of Minkowski space $\mathcal{M}_{5}$. Therefore, the isometry group of the Poincar\'e patch is generated by those Lorentz Killing fields in $\mathcal{M}_{5}$ that are tangential to $\mathcal{N}$. This is a 7-dimensional sub-group of $\Gds$. In the Poincar\'e patch the generators are given by the three space-translations and three 3-rotations on the $\eta={\rm const}$ surfaces and the `dilation' 
\be \label{dilation} D = \frac{1}{l}\, \left[\eta \, \f{\partial}{\partial \eta} + x \f{\partial}{\partial x} + y \f{\partial}{\partial y} + z \f{\partial}{\partial z} \right]. \ee
Therefore, one would anticipate that when $\mathcal{B}^{ab} \hat{=}0$, the asymptotic symmetry group $\G$ of space-times that are \emph{asymptotically de Sitter in a Poincar\'e patch} would not be the full 10-dimensional $\Gds$ but rather this 7-dimensional sub-group thereof. We will now use completeness of $\scri$ to arrive at this result using only the intrinsic structure at $\scri$.

Let us then consider \emph{any} strongly asymptotically de Sitter space-time in which $\scri$ has $\mathbb{R}^{3}$ topology. Since $\scri$ is equipped with a class of conformally flat metrics $q_{ab}$, it is now convenient to work with a flat metric $\qo_{ab}$ in this class and the associated set of Cartesian coordinates $x^{i}$. We can then introduce a convenient basis in the 10-dimensional Lie algebra of $\Gds$: 3 translations $T^{a}$, 3 rotations $R^{a}$, 1 dilation $D^{a}$ and 3 `inverted translations' $\tilde{T}^{a}$ (which are also known as `special conformal transformations'). The dilation is just the extension to $\scri$ of the vector field (\ref{dilation}) and satisfies $\Lie_{D}\, \qo_{ab} = 2 \qo_{ab}$. The inverted translations $\tilde{T}^{a}$ are associated with constant vectors $\To^{a}$ on $(\mathbb{R}^{3}, \qo_{ab})$:
\be \tilde{T}^{a} := 2 (\To_{b}x^{b})\, x^{a} - (x_{b}x^{b})\, \To^{a}\quad {\rm satisfying}\quad  \Lie_{\tilde{T}}\, \qo_{ab} = {4} (\To_{c}x^{c})\, \qo_{ab}, \,\ee
where $x^{a}$ are the Cartesian coordinates of $\qo_{ab}$. The question is if these 10 conformal Killing fields of $\qo_{ab}$ preserve completeness of $\scri$. That is, to begin with, we know that $\scri$ is complete with respect to the given conformal class of metrics $q_{ab}$ (to which $\qo_{ab}$ belongs). Under the 1-parameter family ${\rm d_{V}}(\lambda)$ of diffeomorphisms generated by any of these 10 conformal Killing fields $V^{a}$ we have: $q_{ab} \to q_{ab}(\lambda) = {\rm d}^{\star}_{V}(\lambda)\, q_{ab} = \omega_{\lambda}^{2} q_{ab}$ for an appropriate $\omega_{\lambda}$, and the issue is whether the given manifold $\scri$ continues to be complete with respect to these metrics $q_{ab}(\lambda)$.

To analyze this issue, it suffices to focus just on $\qo_{ab}$. Since $\qo_{ab}$ is left invariant under the action of the 6 Killing fields and rescaled just by a constant, $e^{2\lambda}$, under the action of the dilation, it follows that the diffeomorphisms generated by these 7 symmetry vector fields do preserve completeness of $\scri$. Therefore, we need to examine only the 3 inverted translations $\tilde{T}^{a}$ in some detail. 

Let us begin by recalling the geometric meaning of the inverted translations. Consider the `inverted' coordinates $\tilde{x}^{a} = x^{a}/r^{2}$ which send the origin to the point $i^{o}$ at spatial infinity and $i^{o}$ to the origin. (Here, as usual, $r^{2} = x^{2}+y^{2}+z^{2}$.) Using the $\tilde{x}^{a}$ coordinates we can carry out a one point compactification of $\scri \equiv \mathbb{R}^{3}$ to obtain $\mathbb{S}^{3}$. Coordinates $x^{a}$ cover all of $\mathbb{S}^{3}$ except the `north pole' $i^{o}$ and the coordinates $\tilde{x}^{a}$ cover all of $\mathbb{S}^{3}$ except the south pole (the origin in $x^{a}$ coordinates).  The $\tilde{T}^{a}$ are simply the translations of $\tilde{\qo}_{ab}$; their components are constant in the $\tilde{x}^{a}$ coordinates (which explains the term `inverted translations'). Therefore it is clear that starting from any point (other than the origin) of the original $\mathbb{R}^{3}$ we can reach $i^{o}$ by moving along the integral curves of $\tilde{T}^{a}$ through a finite affine parameter. This suggests that these diffeomorphisms may not preserve completeness of $\scri$.

This is indeed the case. To be specific, let us consider the inverted translation $\tilde{Z}^{a}$ along z-direction (i.e., set $\To^{a}\partial_{a} = \partial/\partial z$). Then, it is easy to verify that the image $\qo_{ab}(\lambda)$ of $\qo_{ab}$ under the  1-parameter family of diffeomorphisms ${\rm d}_{\tilde{Z}}(\lambda)$ is given by 
\be \qo_{ab} (\lambda) = \omega^{2}(\lambda) \qo_{ab} = \f{1}{(1+ 2\lambda z + \lambda^{2}r^{2})^{2}}\,\, \qo_{ab}. \ee
Because $\omega(\lambda)$ goes to zero sufficiently fast as one approaches $i^{o}$, the proper length of, say, the curve $y=0, z=0$ with respect to $\qo_{ab} (\lambda)$ is \emph{finite}, equal to $\pi/\lambda$. Thus, our given 3-manifold $\scri$ is incomplete with respect to $\qo_{ab} (\lambda)$ if $\lambda \not=0$, whence the inverted translation $\tilde{Z}^{a}$ is not a permissible symmetry for our completion that endows $\scri$ with $\mathbb{R}^{3}$ topology. Clearly the same result holds for any inverted translation.

To summarize, as we expected from the conformal completion of the Poincar\'e patch of de Sitter space-time, for conformal completions that make physical space-times asymptotically de Sitter in a Poincar\'e patch, inverted translations are not part of the asymptotic symmetry group $\G$.  This group is now a 7-dimensional sub-group of $\Gds$. If we introduce a basis $T_{i}, R_{i}, D$ with $i =1,2,3$ in the Lie algebra $\g$, then the commutation relations are given by:
\be [D,T_i] =  \frac{1}{l} T_i\, , \quad  [D,R_i]= 0,\quad  [T_i,R_j]
= \epsilon_{ij}{}^k T_k, \quad {\rm and} \quad [R_i,R_j] = \epsilon_{ij}{}^k R_k \, . \label{7dimg}\ee

\emph{Remark:}\\
\indent We have presented the argument for the reduction of $\G$ from 10 to 7 dimensions in some detail because the issue is somewhat subtle. The metric $\qo_{ab}(\lambda)$ is flat because it is just the image of a flat metric $\qo_{ab}$ under a diffeomorphism.  Therefore, each of its Cartesian coordinates can be extended to assume the full range from $-\infty$ to $\infty$. By construction, this extended manifold would be complete w.r.t. $\qo_{ab}(\lambda)$. But this extension is not relevant for us. We are given a conformal completion with $(\scri, \qo_{ab})$ as the boundary. We ask if an inverted translation of $\qo_{ab}$ is an asymptotic symmetry, which requires in particular that the manifold $\scri$ we begin with remain complete with respect to the metrics $\qo_{ab}(\lambda)$. What we showed is that it does not. The discussion of isometries of the Poincar\'e patch and of the geometrical meaning of inverted translation is, strictly speaking, not needed for the final result. But it brings out the reason as to why certain symmetries cease to be permissible as we move from $\mathbb{S}^{3}$ topology to $\mathbb{R}^{3}$ topology.

\subsubsection{$\mathbb{S}^{2}\times \mathbb{R}$ topology}
\label{s4.3.3}

The $\mathbb{S}^{2}\times \mathbb{R}$ topology results when one removes a point ---say, the origin--- from $\mathbb{R}^{3}$. The 3 translations $T^{a}$, that are permissible in the case when the topology is $\mathbb{R}^{3}$, fail to leave the origin invariant. Alternatively, starting from a point on $\mathbb{S}^{2}\times \mathbb{R}$ one can reach the origin in $\mathbb{R}^{3}$ by moving a finite affine parameter distance along the integral curve of one of the 3 translations. Therefore, one would expect that the same reason that made the inverted translations $\tilde{T}^{a}$ inadmissible as asymptotic symmetries in the $\mathbb{R}^{3}$ case, would now make the three translations $T^{a}$ inadmissible. This is indeed the case. We will now show that the symmetry group of $\scri$ with topology $\mathbb{S}^{2}\times \mathbb{R}$ is generated only by the three rotations and the dilation.

Let us begin with the 3-metric $q_{ab}$ on $\scri$ of the Schwarzschild-de Sitter space-time discussed in section \ref{s3.2}:
\be  q_{ab} {\rm d} x^a {\rm d} x^b\, =\, {\rm d}t^{2} + l^{2}\, d\omega_{2}^{2} \, =\, {\rm d}t^{2} + l^{2}\,({\rm d} \theta^2 + \sin^2 \theta \, {\rm d} \phi^2) \ee 
for which $\scri$ is complete. It can be cast in a manifestly conformally flat form by introducing $r := l\, \exp (t/l)$:
\be  q_{ab} {\rm d} x^a {\rm d} x^b\, =\, \f{l^{2}}{r^{2}}\,\,\big({\rm d}r^{2} + r^{2}\, {\rm d}\omega_{2}^{2} \big)\, =\, \f{l^{2}}{r^{2}} \,      {\mathring{q}}_{ab} {\rm d} x^a {\rm d} x^b\,  \, . 
\ee
As expected, the origin of the flat metric $\qo_{ab}$ is not in $\scri$ because $q_{ab}$ is ill-defined there. Under the 1-parameter family of diffeomorphisms generated by, say, the z-directional translation $Z^{a}$, the image $d^{\star}_{Z}(\lambda)\, q_{ab}\, =:\, q_{ab}(\lambda)$ of $q_{ab}$ is given by:
\be q_{ab}(\lambda) = \f{l^{2}}{x^{2}+y^{2}+ (z+\lambda)^{2}}\,\, \qo_{ab} \ee
whence the distance between the point, say, $x=0,y=0, z=z_{o}$ and the origin with respect to $q_{ab}(\lambda)$ is \emph{finite}, given by $l\, \ln ((z_{o}+\lambda)/\lambda)$. Thus, when $\lambda \not= 0$, the manifold $\scri$ we began with ---which is complete with respect to the metric $q_{ab}$--- is no longer complete. Therefore the translations $T^{a}$ are not permissible symmetries, whence the symmetry group of asymptotically Schwarzschild-de Sitter space-times is only 4-dimensional, generated by the three rotations $R_{i}$ and the dilation $D$, with commutation relations given by (\ref{7dimg}). Now, there is a preferred time translation, represented by $D$.%
\footnote{Strictly, we should also check that the inverted translations of $\qo_{ab}$ are also not symmetries but the argument is essentially the same as that in section \ref{s4.3.2}.}\\

In summary, when the leading order term of the magnetic part of the Weyl tensor $\mathcal{B}^{ab}$ is set to zero, the intrinsic metric on $\scri$ is conformally flat. This leads to a drastic reduction of the group of asymptotic symmetries from the infinite dimensional $\Diff$ to a finite dimensional group. Had we not included the completeness requirement in Definition 2, then the universal structure would have included \emph{all} conformally flat metrics and the symmetry group would have been the de Sitter group. However, completeness requirement is important, e.g., to have a useful notion of black holes.%
\footnote{A space-time $(M, g_{ab})$ admits a black-hole region $B$ if 
the past of $\scri^{+}$ does not cover $M$. Therefore if we were to drop the completeness requirement, one could just attach a portion of full $\scri^{+}$ to, say, the de Sitter space-time, consider its past, and arrive at the absurd conclusion that de Sitter space-time contains a black hole! For the same reason, $\scri^{+}$ is required to be complete in the asymptotically Minkowski case as well \cite{gh}.} 
With completeness requirement, only a sub-class of conformally flat metrics is permissible when the topology of $\scri$ is either $\mathbb{R}^{3}$ or $\mathbb{S}^{2}\times \mathbb{R}$. The asymptotic symmetry group $\G$ is then a proper sub-group of the de Sitter group $\Gds$: it is 7-dimensional for $\mathbb{R}^{3}$ topology and 4-dimensional for 
$\mathbb{S}^{2}\times \mathbb{R}$ topology. In applications, the $\mathbb{S}^{3}$ topology is useful primarily for studying perturbations of pure de Sitter space-times. In the cosmological context we are led to the $\mathbb{R}^{3}$ case and for isolated gravitating systems to the $\mathbb{S}^{2}\times \mathbb{R}$ case. Therefore the corresponding symmetry groups will play an important role in subsequent papers. This interplay between topology and asymptotic symmetries is a new feature associated with the $\Lambda >0$ case; it does not arise in the asymptotically Minkowski context where the topology of $\scri$ is always $\mathbb{S}^{2}\times \mathbb{R}$ and, in presence of gravitational radiation, the asymptotic symmetric group is always the BMS group $\B$.

\section{The $\mathcal{B}^{ab}\, \hat{=} \,0$ condition}
\label{s5}

In this section we will probe the geometrical and physical meaning of the stronger boundary condition $\mathcal{B}^{ab}\, \hat{=} \, 0$ at $\scri$ and show that it is a severe restriction which cannot be justified, or even motivated, on physical grounds. Since this is an important issue, we will proceed in three steps. 

We will begin in section \ref{s5.1} with the $\Lambda=0$ case and analyze the implication of the $\mathcal{B}^{ab}\, \hat{=} \, 0$ condition at $\scri$ in the well-understood asymptotically Minkowski context. The additional restriction now reduces the BMS group to the 10-dimensional Poincar\'e group, just as it reduced $\Diff$ to the 10-dimensional de Sitter group in section \ref{s4.3.1}. However, we will find that the condition implies that there is \emph{no} gravitational radiation at $\scri$ ! Given the close parallel between the reductions of symmetry groups in the two cases, the last result can be taken to be an indication that the condition $\mathcal{B}^{ab} \, \hat{=} \, 0$ is inadmissible in the $\Lambda>0$ case as well. But recall that since $\scri$ is null in the $\Lambda= 0$ case, the electric and magnetic parts of the Weyl tensor are not independent. Therefore the question naturally arises: Is this implication of absence of gravitational radiation in the $\Lambda=0$ case perhaps tied with the fact that $\scri$ is null in this case? 

To probe this issue, in section \ref{s5.2} we examine test Yang-Mills fields in de Sitter space-time \emph{where $\scri$ is space-like} and analyze the implications of the analogous additional condition $B^{a}_{i} \, \hat{=} \,  0$ in the Yang-Mills sector. This analysis enables one to separate effects that can be attributed primarily to the null nature of $\scri$ in the $\Lambda=0$ case from those that originate from the non-Abelian character of the interaction ---i.e., the fact that in the Yang-Mills theory (and general relativity), fields act as their own source. We will show that the condition $B^{a}_{i} \, \hat{=} \,  0$ does reduce the local gauge group at $\scri$ to the global gauge group, just as $\mathcal{B}^{ab} \, \hat{=} \, 0$ reduces $\Diff$ to $\Gds$ in the gravitational case. This enables one to define Yang-Mills charges unambiguously. However, they are absolutely conserved; even though Yang-Mills fields are sources of their own charges, there is no leakage through $\scri$. Furthermore, while a restricted class of Yang-Mills waves is permissible at $\scri$,  they do not carry de Sitter energy-momentum or angular momentum even locally. Thus the requirement $\mathcal{B}^{a}_{i} \, \hat{=} \, 0$ is again a strong restriction that cannot be justified.

Finally, in section \ref{s5.3}, we consider the full, non-linear gravitational field in the $\Lambda>0$ case. The discussion of the first two sub-sections provides a deeper understanding of the structures at play in this case. We will find that the technical implication of the condition $\mathcal{B}^{ab}\hat{=} 0$ is closer to that of $B^{a}_{i} \hat{=}0$ in the Yang-Mills case: while it does not completely rule out gravitational waves at $\scri$, it is a severe restriction because it implies that the local fluxes of energy-momentum and angular momentum carried by gravitational waves across $\scri$ must all vanish.  

\subsection{Asymptotically Minkowski space-times}
\label{s5.1}

Fix an asymptotically Minkowski space-time $(\hat{M}, \hat{g}_{ab})$ and a completion $(M, g_{ab})$ thereof with a conformal factor $\Omega$. Then, $\scri$ is topologically $\mathbb{S}^{2}\times \mathbb{R}$ and, as discussed in section \ref{s4.1}, endowed with a pair of fields $(q_{ab}, n^{a})$. This leading order structure is common to all asymptotically Minkowski space-times. The next order structure is the pull-back $D$ of the space-time connection $\nabla$ compatible with $g_{ab}$. The pull-back is well defined because $\nabla_{a} n_{b} \, \hat{=} \,  0$ in the divergence-free conformal frames that are used, and satisfies $D_{a} q_{bc} \, \hat{=} \,  0$ and $D_{a}n^{b} \, \hat{=} \, 0$. However, $D$ is not uniquely determined by these properties because it is degenerate. What physical information does it encode? Recall first that there is still considerable restricted conformal freedom. Therefore one is naturally led to consider equivalence classes $\{D\}$ of conformally related derivative operators $D$. The difference between the $\{D\}$ arising from any two space-times is characterized by a trace-free symmetric tensor $\sigma_{ab}$ defined intrinsically on $\scri$ which is transverse to $n^{a}$ (in the sense that $\sigma_{ab}n^{b} \, \hat{=} \, 0$). The two independent components of $\sigma_{ab}$ correspond to the two radiative modes of the gravitational field in full, non-linear general relativity. (For proofs of results quoted in this sub-section, see \cite{aa-asym}.)

Since $\scri$ is 3-dimensional, the Riemann curvature $\ub{R}_{abc}{}^{d}$ of $D$ is completely captured in a second rank tensor $\ub{S}_{a}{}^{b}$:
\be 
\ub{R}_{abc}{}^{d} \, \hat{=} \,  \frac{1}{2}(q_{c[a} \ub{S}_{b]}{}^d - \ub{S}_{c[a} \delta_{b]}{}^d)
\ee
where $\ub{S}_{ab} = \ub{S}_{a}{}^{c}\, q_{bc}$. Under (the restricted) conformal transformations, $\ub{S}_{a}{}^{b}$ has a complicated transformation property. However, using the fact that the space of integral curves of $n^{a}$ is topologically $\mathbb{S}^{2}$, it is possible to extract the conformally invariant information in $\ub{S}_{a}{}^{b}$. This is encoded in two tensor fields at $\scri$. The first is a symmetric, transverse, trace-free tensor $N_{ab}$, called the \emph{Bondi news tensor} \cite{rg}. A common strategy is to use a `Bondi conformal frame' in which ($n^{a}$ is divergence-free and in addition) $q_{ab}$ is the unit 2-sphere metric. In a Bondi frame, $N_{ab}$ is simply the trace-free part of $\ub{S}_{ab}$. Fluxes of \emph{all} the BMS momenta carried away by gravitational waves all vanish if $N_{ab}=0$. The second tensor field is the `magnetic part' $\mathcal{B}^{ac}$ of the asymptotic Weyl curvature:
\be \mathcal{B}^{ac} = {}^{\star}\!K^{abcd}\, n_{b}\, n_{d}
\quad {\rm where} \quad K^{abcd} = \lim_{\rightarrow\scri}\,\,
 \Omega^{-1}\, C^{abcd}\, . \ee
The two tensor fields are related by 
\be \mathcal{B}^{ab} = 2\, \epsilon^{amn}\, D_{m} N_{n}{}^{b} \ee

Now let us consider space-times for which  $\mathcal{B}^{ab}\,\hat{=} \,0$.  Then we have $D_{[m}\,N_{n]p}\, \hat{=}\, 0$. Transvecting this equation with $n^{a}$ and using $n^{a}N_{ab}\, \hat{=}\, 0$ and $D_{a}n^{b}\, \hat{=} \,0$, one obtains $\Lie_{n}\, N_{ab}\, \hat{=} \,0$. Therefore $N_{ab}$ admits an unambiguous projection to the 2-sphere $S$ of generators of $\scri$, which we denote by $\bar{N}_{ab}$. Without loss of generality one can work in a Bondi frame. Then, on a unit 2-sphere $S$, we have a tensor field $\bar{N}_{ab}$ satisfying
\be \bar{N}_{[ab]}=0,\quad  \bar{N}_{ab} \bar{q}^{ab} =0, \quad {\rm and} \quad \bar{D}_{[a} \, \bar{N}_{b]c} =0\, , \ee
where $\bar{q}_{ab}$ is the unit 2-sphere metric on $S$ induced by $q_{ab}$ on $\scri$, and $\bar{D}$ is its torsion-free derivative operator. Appendix A shows that the only solution to these equations is $\bar{N}_{ab} =0$ on $S$ which immediately implies $N_{ab} \hat{=} 0$ on $\scri$.%
\footnote{Since $\Lie_{n}\, N_{ab} \hat{=}0$, if $N_{ab}$ were not to vanish and $\scri$ were to be complete, then the total flux of Bondi energy over $\scri$ would be infinite. Therefore invoking the obvious physical considerations one can conclude that $N_{ab}$ must vanish on $\scri$. Appendix A establishes this result without assuming completeness of $\scri$. Also an intermediate result in the proof holds in higher dimensions and has been used in different contexts, including the analysis of the structure at spatial infinity \cite{aarh,spatial-review,asymflat-others}.} 
Therefore, if $\mathcal{B}^{ab}\, \hat{=}\,0$ on $\scri$, connections $\{D\}$ have trivial curvature. 

In space-times with $\mathcal{B}^{ab} =0$ at $\scri$, it is then natural to include the `trivial' equivalence class of connections $\{\mathring{D}\}$ in the list of universal structures at $\scri$. For this family of space-times, the asymptotic symmetry group is the sub-group of the BMS group $\B$ that also leaves this  $\{\mathring{D}\}$ invariant. It turns out that this reduces the infinite dimensional BMS group $\B$ to a 10-dimensional Poincar\'e sub-group thereof. This consequence of the $\mathcal{B}^{ab} =0$ condition in the $\Lambda=0$ case is completely analogous to that in the $\Lambda>0$ case. However, since $\mathcal{B}^{ab} =0$ implies that the Bondi news tensor $N_{ab}$ must also vanish, these space-times \emph{admit no gravitational waves.} Therefore, outside the limited context of stationary space-times, the requirement is extremely strong.\\

\emph{Remark}:\\
The definition $\mathcal{B}^{ac}\, \hat{=} \,{}^{\star}\!K^{abcd}n_{b}n_{d}$ of the magnetic part of the Weyl tensor used here is the natural analog of that used in the decomposition of the Weyl tensor on space-like and time-like surfaces.  In the Newman-Penrose notation, $\mathcal{B}^{ab} \, \hat{=} \,  0$ is equivalent to $\Psi^{0}_{4}\, \hat{=} \, 0,\, \Psi^{0}_{3}\, \hat{=} \, 0$ and ${\rm Im} \Psi_{2}^{0} \, \hat{=} \,  0$. Therefore, it is not surprising that $\mathcal{B}^{ab} \, \hat{=} \,  0$ implies that the Bondi news $N_{ab}$ vanishes. However, because the normal $n^{a}$ is now also tangential to $\scri$ when it is null, a key difference arises: the electric and magnetic parts are no longer independent. The five components of $\mathcal{E}^{ab}$ correspond to $\Psi^{0}_{4},\, \Psi^{0}_{3}$ and ${\rm Re} \Psi_{2}^{0}$. Thus, the condition $\mathcal{B}^{ab}\, \hat{=} \, 0$ imposes a strong condition on $\mathcal{E}^{ab}$ as well; now its only non-zero component is ${\rm Re} \Psi_{2}^{0}$. This determines the Bondi mass which is in general non-zero but absolutely conserved because there is no gravitational radiation to carry away energy-momentum.

\subsection{Yang-Mills fields in de Sitter space-time}
\label{s5.2}

Let us now consider source-free Yang-Mills fields $\hat{F}_{ab}^{i}$ in the de Sitter space-time $(\hat{M}, \hat{g}_{ab})$, where the index $i$ refers to the Lie algebra of the internal group which we will take to be an $n$-dimensional compact group $G$. Since the Yang-Mills equation is conformally invariant, if we denote the Yang-Mills connection  by $\hat{A}_{a}^{i}$, then $A_{a}^{i} = \hat{A}_{a}^{i}$ satisfies the Yang-Mills equation on the conformally rescaled space-time $(M,g_{ab})$. We assume that $A_{a}^{i}$ admits a smooth extension to $\scri$ (which is compatible with the requirement on the stress-energy tensor in Definition 1). 

At $\scri$, we can define two types of physical quantities for Yang-Mills fields. The first of these are associated with isometries of de Sitter space-time: given a Killing field $\xi^{a}$, 
\begin{align} \F_{\xi}[\Delta\scri] &:=  \int_{\Delta \scri}\, T_{ab}\, \xi^{a} \no^{b}\,\, {\rm d}^{3}V  \nonumber\\
&= -\frac{1}{4 \pi} \, \int_{\Delta \scri}\, \epsilon_{abc}\, E^{a}_{i}\, B^{b}{}^{i} \,\xi^{c}\,\,  {\rm d}^{3}V \label{YMfluxes}
\end{align}
defines the flux of the `de Sitter momentum' across any patch $\Delta\scri$ of $\scri$. Here $\no^{a}$ is the \emph{unit} normal to $\scri$, $T_{ab}$ denotes the stress-energy tensor of the Yang-Mills fields, $E^{a}_{i} =  F^{ab}_{i}\, \no_{b}$,\, $B^{a}_{i} = {}^{\star}\!F^{ab}_{i} \,\no_{b}$, and in the second step we have used the fact that every Killing field $\xi^{a}$ is tangential to $\scri$. The second type of physical quantity arises from the fact that, because of its non-Abelian nature, the Yang-Mills field carries its own charge. Therefore, given any 2-sphere $S$ and a Lie-algebra-valued field $\zeta^{i}$ on it, we can define electric and magnetic type charges:
\be Q[S,\zeta] := \f{1}{4\pi}\, \oint_{S}\, E^{a}_{i}\,\zeta^{i}\, dS_{a}, \quad  {\rm and} \quad Q^{\star}[S,\zeta] := \f{1}{4\pi}\, \oint_{S}\, B^{a}_{i}\,\zeta^{i}\, dS_{a}. \label{YMcharges}\ee
However, because of the local gauge freedom in choosing the generator $\zeta^{i}$, in each class we have an infinite family of charges. If we vary $S$, the values of these charges change because of two reasons. First, there is a leakage of the Yang-Mills fields between any two 2-spheres and, second, the generators $\zeta^{i}$ can change in an arbitrary fashion from one 2-sphere to another.

To compare charges in a useful manner one needs a rigid prescription to choose $\zeta^{i}$ on any 2-sphere $S$ to enable one to say that one is comparing the values of the `same' charge on two different 2-spheres. This can be achieved by restricting oneself only to those Yang-Mills fields for which $B^{a}_{i}$ vanishes at $\scri$. Then the pull-back $\pb{F}_{ab}^{i}$ to $\scri$ of the Yang-Mills field vanishes and we can introduce \emph{covariantly constant} Lie-algebra valued fields $\zeta^{i}$. Thus, the gauge group is reduced from the infinite dimensional group ${\rm Loc}[G]$ of local gauge transformations to the n-dimensional group $G$ of global gauge transformations, just as the infinite dimensional $\Diff$ is reduced to the 10-dimensional $\Gds$ by imposing $\mathcal{B}^{ab} \, \hat{=} \,  0$. Furthermore, using these covariantly constant $\zeta^{i}$, we can define $n$ charges $Q_{\zeta}[S]$ on any 2-sphere, associated with the global gauge group. Therefore, it is now  meaningful to compare charges on two different 2-spheres $S_{1}$ and $S_{2}$ and attribute the difference solely to the leakage of the Yang-Mills fields between $S_{1}$ and $S_{2}$. In view of this nice structure, it is tempting to regard the additional condition $B^{a}_{i} \,\hat{=} \,0$ as a natural restriction.

However, from (\ref{YMfluxes}) we see that if $B^{a}_{i} \, \hat{=} \,  0$, then \emph{all 10 de Sitter fluxes vanish identically}. Furthermore, this happens across \emph{any} local region $\Delta\scri$ of $\scri$; i.e., not because of a subtle cancellation between different regions. Secondly, while now we have well-defined global charges $Q_{\zeta}[S]$, because we can now choose a gauge such that $A_{a}^{i} =0$ in the region enclosed by any given two 2-spheres $S_{1}$ and $S_{2}$, we have
\begin{align} Q_{\zeta}[S_{1}]\, -\, Q_{\zeta}[S_{2}] 
 &= \int_{\Delta\scri}\, (D_{a}E^{a}_{i}) \,\zeta^{i} \, {\rm d}^{3}V\nonumber\\
&= \int_{\Delta\scri}\, {\rm Div}_{A}(E_{i})\, \zeta^{i}\, {\rm d}^{3} V =0 \end{align}
where $\Delta\scri$ is the region of $\scri$ bounded by $S_{1}$ and $S_{2}$,\,\, ${\rm Div}_{A}$ is the gauge covariant divergence, and where in the last step we have used the Gauss law. Thus, the charges $Q_{\zeta}$ are independent of the choice of the 2-sphere $S$. Finally, since $\scri$ is topologically $\mathbb{S}^{2}$, one can just contract any 2-sphere $S$ indefinitely, whence $Q_{\zeta}[S] =0$ for charges and all $S$. 

The vanishing of the fluxes and charges already shows that the restriction $B^{a}_{i} \, \hat{=} \,  0$ is quite severe from a physical viewpoint. Mathematically, we can just refer to the Cauchy problem at $\scri$ to conclude that the restriction \emph{removes half the degrees of freedom.} We can still have transverse electric fields $E^{a}_{i}$ at $\scri$ but they will have zero local fluxes of de Sitter energy, momentum and angular momentum and all the electric charges will also vanish.\\

\emph{Remarks:}

1) Had we worked in the Poincar\'e patch of de Sitter space-time, $\scri$ would be topologically $\mathbb{R}^{3}$. We could also have considered space-times which are asymptotically Schwarzschild-de Sitter where $\scri$ is topologically $\mathbb{S}^{2}\times \mathbb{R}$. In both cases, the same analysis shows that all the local fluxes (\ref{YMfluxes}) vanish and the n electric charges (\ref{YMcharges}) are independent of the surface $S$. However, in the $\mathbb{S}^{2}\times \mathbb{R}$ case, these charges need not be zero. This is completely analogous to what happens in stationary space-times in the asymptotically Minkowski context where the total 4-momentum and angular momentum of the space-time can be non-zero but fluxes of these `charges' across any patch of $\scri$ vanish identically.

2) In the Yang-Mills case, charges (\ref{YMcharges}) refer to the \emph{internal} group while the fluxes (\ref{YMfluxes}) refer to the asymptotic \emph{space-time} symmetries. In the gravitational case the charges are again 2-sphere integrals and fluxes are 3-surface integrals. But now they both refer to the asymptotic space-time symmetries and are therefore intertwined: fluxes account for the differences between the gravitational charges associated with different 2-spheres.

\subsection{Non-linear gravitational fields with $\Lambda >0$}
\label{s5.3}

Let us now consider strongly asymptotically de Sitter space-times. For simplicity of presentation, we will first consider the case when $\scri$ is topologically $\mathbb{S}^{3}$ and then discuss the other, more interesting topologies.  

\subsubsection{Further consequences of field equations and Bianchi identities}
\label{s5.3.1}

Since $\mathcal{B}^{ab}\, \hat{=} \, 0$ and $\scri$ has $\mathbb{S}^{3}$ topology, it is equipped with a 10-dimensional group of conformal isometries, isomorphic to $\Gds$. The natural question is whether, given a 2-sphere $S$ on $\scri$, we can associate with each generator $\xi^{a}$ of these isometries a gravitational charge $Q_{\xi}[S]$ and analyze the fluxes $\mathcal{F}_{\xi} [\Delta\scri]$ across regions $\Delta\scri$ bounded by two 2-spheres. These would be the analogs of the Bondi charges \cite{td} and fluxes \cite{aams} in asymptotically Minkowski space-times discussed in section \ref{s5.1} and the charges and fluxes of Yang-Mills fields discussed in section \ref{s5.2}. To probe this issue we need to find suitable consequences of the field equations and Bianchi identities that would motivate the definitions of charges and lead to the balance laws in terms of appropriate fluxes. This discussion will be parallel to that in the asymptotically anti-de Sitter case \cite{aaam} once the difference in the sign of $\Lambda$ has been taken into account. However, there is a subtle error in the intermediate steps of that discussion which led to the omission of 
terms involving the trace of the stress energy tensor of matter in the final result. We will take this opportunity to correct that error.

Let us begin by considering the contracted Bianchi identity to the next leading order from that considered in (\ref{eqcontractedbianchi}). Set
\be \ub{T}_{a}{}^{b} :=  \Omega^{-3}\, \hat{T}_{a}{}^{b}\ee
which has a smooth limit to $\scri$ and treat it as a tensor field in the conformally completed space-time, whose indices are raised and lowered with the rescaled metric $g^{ac}$ and $g_{cb}$. Then on $\hat{M}$ we have the identity:
\be \label{identity}
\nabla^{m}\, K_{abcm} =  \frac{8\pi\, G}{\Omega}\, \Big[- 2 n_{[a}\, \ub{T}_{b]}{}^{m}g_{mc} + \ub{T} n_{[a} g_{b]c} + g_{c[a} \ub{T}_{b]}{}^{m}n_{m} + \frac{1}{3} \Omega (\nabla_{[a} \ub{T}) g_{b]c} - \Omega \nabla_{[a} \ub{T}_{b]}{}^{m}g_{mc}\Big] . \ee
Transvecting this equation with $n^{a}n^{c}$, projecting the free index by $q^{bp}$ and taking the limit to $\scri$ we obtain:
\be \label{divE}
D_{m}\mathcal{E}^{mp}\, \hat{=} \, \lim_{\to \scri}\,\, 4\pi\, G\,\Big[ (1/l)\, J^{p} +  \frac{1}{3} D^{p} \ub{T} + n^{a} \nabla_{a} \big(\ub{T}_{bc}\, q^{bp} (\frac{n^{c}}{n\cdot n})\big) - D^{p}\big(\ub{T}_{ac} \, (\frac{n^{a}n^{c}}{n\cdot n})\big)\Big] \, . \ee
Here 
\be \label{J}  \mathcal{E}^{mp}\, \hat{=} \, \lim_{\to\scri}\, K^{ampb}\, \mathring{n}_{a} \mathring{n}_{b}, \quad {\rm and} \quad
J^{p} \, \hat{=} \, \lim_{\to\scri}\, \Omega^{-4}\, \hat{T}_{a}{}^{b} q^{ap}\, \mathring{n}_{b}\ee
and $\mathring{n}^{a}\, \hat{=}\, l\,n^{a}$ is the unit (future pointing) normal to $\scri$. $\mathcal{E}^{mp}$ is the leading order, electric part of Weyl curvature, and  $J^{p}$ is the leading order matter current at $\scri$. (Recall that the limit to $\scri$ of $\Omega^{-3}\, \hat{T}_{a}{}^{b} q^{ap}\, n_{b}$  vanishes.) Eq. (\ref{divE})  simplifies considerably on using again the fall-off conditions (\ref{falloff}) which tell us that only the component $\ub{T}_{ab}n^{a}n^{b}$ of $\ub{T}_{ab}$ can be non-zero at $\scri$. We obtain:
\be \label{conservation}
D_{m}\mathcal{E}^{mp} \, \hat{=}\, {8\pi \, G}\,\,\big(\f{1}{l} J^{p} - \f{1}{3}\, D^{p} \ub{T}\big) \, .  \ee 
This will serve as the key equation for defining charges and fluxes at $\scri$. \\

\emph{Remark:}\\
Eqs. (\ref{identity}), (\ref{divE}) and (\ref{conservation}) are conformally covariant, as they must be: under $g_{ab} \to g^{\prime}_{ab}= \omega^{2}g_{ab}$ (with $n^{a} \nabla_{a} \omega \, \hat{=}\, 0$), they just get multiplied by a power of $\omega$. However, explicit checks of this behavior is rather subtle for the last two of these equations because various components of $\ub{T}_{a}^{b}$ have different asymptotic behavior.

\subsubsection{Gravitational Charges}
\label{s5.3.2}

We can now obtain appropriate balance laws starting from Eq. (\ref{conservation}). Recall that if $\mathcal{B}^{ab}\, \hat{=} \, 0$, infinitesimal asymptotic symmetries are represented by conformal Killing fields on $\scri$. Let us transvect (\ref{conservation}) with a conformal Killing field $\xi^{a}$ and integrate over a portion of $\scri$ bounded by two 2-spheres (or, more generally, any two 2 compact surfaces). By integrating by parts and using the fact that $\mathcal{E}^{ab}$ is symmetric and trace-free, we obtain:
\be \label{balance1}
\f{l}{8\pi G}\Big(\oint_{S_{2}}\, -\, \oint_{S_{1}}\Big) (\mathcal{E}_{ab}\, + \, \frac{8\pi G}{3}\, \ub{T} q_{ab})\,\xi^{a} dS^{b} = \int_{\Delta\scri} (J_{a}\xi^{a} \, + \, \alpha \ub{T} l)\, \rmd^{3}V   \ee
where the function $\alpha$ is given by
\be \mathcal{L}_{\xi}\, q_{ab}\, = \, 2\alpha\, q_{ab}\, .  \ee
At first glance, the first term on the right side of (\ref{balance1}) would appear to be the matter flux associated with $\xi^{a}$ through the region $\Delta\scri$. However, it is not conformally covariant: under $g_{ab} \to g^{\prime}_{ab}\, = \,\omega^{2} g_{ab}$  (with $n^{a}\nabla_{a} \omega \, \hat{=}\, 0$), on $\scri$ we have:
\be {J^{\prime}}^{p}\, \xi^{\prime}_{p}\, \hat{=} \, \omega^{-3} \left(J^{p}\xi_{p} - (\omega^{-1}\mathcal{L}_{\xi}\, \omega)\, l\, \ub{T}\right)\, . \ee
The second term transforms as 
\be  \alpha^{\prime}\ub{T}^{\prime} l\, \hat{=} \,\omega^{-3}\, \left(\alpha + \omega^{-1}\, \mathcal{L}_{\xi}\omega\right)\,\left(\ub{T}\, l\right)\, . \ee
Therefore the sum of the two terms on the right of (\ref{balance1}) is conformally covariant and, since $\rmd^{3}V^{\prime} =\omega^{3}\rmd^{3}V$, the integral on the right is conformally \emph{invariant.} Thus, to define a physically viable matter current we need the second term on the right side; it is the sum that represents $\mathcal{F}_{\xi}^{\rm{matt}}\, [\Delta\scri]$, the flux of the $\xi^{a}$ component of the de Sitter momentum across $\Delta \scri$, carried by matter.

In view of (\ref{balance1}) we are now led to define the 2-sphere integrals on the left hand side as charges $Q_{\xi}[S]$ associated with the generator $\xi^{a}$ of the asymptotic symmetry. Therefore we set:
\be Q_{\xi}[S] \, :=\, - \f{l}{8\pi G}\, \oint_{S} \left(\mathcal{E}^{ab}\, +\, \frac{8\pi G}{3}\, \ub{T}\, q^{ab}\right)\, \xi_{a} \mathring{r}_{b}\, \rmd^{2}V  \label{gravcharges}\ee 
where $\mathring{r}^{a}$ is the unit normal to $S$ and $\rmd^{2}V$ the volume element on $S$. This is the gravitational charge associated with the asymptotic symmetry generator $\xi^{a}$ and the 2-surface $S$. Again, for these charges to have physical content, they should refer only to the physical space-time under consideration and not to the choice of the conformal completion made in their definition. Under a rescaling  we have: 
\be \mathcal{E}^{\prime}_{ab}\,  \hat{=} \, \omega^{-1}\, \mathcal{E}_{ab},\quad \ub{T}^{\prime}\, q^{\prime}_{ab} = \omega^{-1} \ub{T}\, q_{ab},\quad {\xi^\prime}^{a} = \xi^{a}, \quad \mathring{r^{\prime}}^{b}\, \hat{=} \,  \omega^{-1}\mathring{r}^{b}\quad {\rm and} \quad \rmd^{2}V^{\prime}\, \hat{=}\, \omega^{2}\, \rmd^{2}V\, .\ee
Therefore the charges (\ref{gravcharges}) are indeed conformally invariant. (Similarly, the flux integral on the right side of (\ref{balance1}) is also conformally invariant, as it must be for consistency.) Thus, the charge integrals are indeed insensitive to the choice of conformal completion.

The charge integrals (\ref{gravcharges}) and the balance laws (\ref{balance1}) have two novel features from the viewpoint of asymptotically Minkowski space-times. We will conclude this sub-section with a discussion of these differences.

The first feature is the appearance of a matter term in the integrand of the gravitational charge integral.  However, recall that in the asymptotically Minkowski space-times, one asks for a stronger fall-off for the stress energy tensor, namely that $\Omega^{-2}\hat{T}_{ab}$ should have a limit to $\scri$ (rather than $\Omega^{-1}\hat{T}_{ab}$ as in Definition 1). Had we imposed this stronger condition, then the field $\ub{T}$ would vanish on $\scri$ and the extra term would disappear. As discussed in section \ref{s2.1}, the stronger condition is in fact natural for Yang-Mills fields and null fluids and the gravitational charge is then expressed entirely in terms of geometry. However, as we also pointed out in section \ref{s3.5}, the weaker condition is necessary to accommodate cosmological solutions. In the Friedman-Lema\^itre cosmologies, it turns out that the extra term integrates out to zero for all conformal Killing fields in the charge integrals as well as balance laws if we restrict ourselves to round 2-spheres. However, in the general cosmological contexts, such as those considered in \cite{hr}, they would play an interesting role. 

The second difference from the asymptotically Minkowski case is more significant. In absence of matter sources, charges (\ref{gravcharges}) are \emph{absolutely conserved,} i.e, do not depend on the choice of the 2-sphere $S$ used in their evaluation. Therefore, if $\scri$ is topologically $\mathbb{S}^{3}$ or $\mathbb{R}^{3}$, we can just continuously shrink the 2-sphere to a point to conclude $Q_{\xi}[S] =0$ for any asymptotic symmetry generator $\xi^{a}$ and any 2-sphere $S$. In particular, as one would expect physically, all gravitational charges vanish identically in de Sitter space-time.  Interesting cases correspond to isolated systems where the topology of $\scri$ is $\mathbb{S}^{2}\times \mathbb{R}$ and the asymptotic symmetry group is 4-dimensional, with one time translation (or dilation) and three rotations.

\subsubsection{Examples}
\label{s5.3.3}

A first viability test of the definition (\ref{gravcharges}) is provided by computing these charges in simple examples. In the Schwarzschild-de Sitter space-time, using the conformal completion discussed in section \ref{s3.2}, we obtain:
\be \mathcal{E}_{ab} \, \hat{=} \, -\f{3GM}{l^{3}}\, \big(D_{a}t D_{b}t - \f{1}{3} q_{ab}\big).  \ee
Since the matter current $J^{a}$ and $\ub{T}$ vanish identically, the charge integrals are independent of the choice of the 2-sphere $S$. Therefore we can evaluate them on round 2-spheres. A simple calculation gives:
\be Q_{t}[S] = M \quad {\rm and} \quad Q_{R}[S] = 0, \ee
where $t^{a}$ is the time translation, $t^{a}\partial_{a} = \partial/\partial t$  and $R^{a}$ any rotational Killing vector. 

Next, consider the conformal completion of the Kerr space-time discussed in section \ref{s3.2}. Now, the expressions are much more involved:
\begin{align}
\mathcal{E}_{ab}\,  &\hat{=} - \frac{GM}{l (1+ \frac{a^2}{l^2})^2} \Big( \frac{2}{l^2} \grad_a t \grad_b t - \frac{\left(1+ \frac{a^2}{l^2}\right)^2}{1+ \frac{a^2}{l^2} \cos^2 \theta} \grad_a \theta \grad_b \theta  \\
& \qquad - \sin^2 \theta \big[1 - \frac{a^2}{l^2} \big(\frac{1}{2} - \frac{3}{2} \cos 2\theta \big) \big] \grad_a \phi \grad_b \phi - \frac{4a}{l^2} \sin^2 \theta \grad_{(a} t \grad_{b)} \phi \Big)\,.
\end{align}
The unit normal $\mathring{r}^{a}$ to the 2-spheres, $t =$const, and the intrinsic volume element ${\rmd}^{2}V$ on them are given by:
\begin{align}
\mathring{r}^a\partial_{a}\, & \hat{=} \, \sqrt{\frac{1+\frac{a^2}{l^2}}{1+ \frac{a^2}{l^2}\cos^2\theta}} \left( \left(1+\frac{a^2}{l^2} \right) \left(\frac{\partial}{\partial t}\right) + \frac{a}{l^2}\left(\frac{\partial}{\partial \phi} \right)\right)\, , \\
{\rmd}^{2}V\, & \hat{=}\, \frac{l^2 \sin \theta}{\sqrt{1 + \frac{a^2}{l^2} \cos^2 \theta} \sqrt{1+\frac{a^2}{l^2}}}\, {\rm d}\theta\, {\rm d}\phi\, .
\end{align}
A direct evaluation of the charge integrals $Q_{\xi}[S]$ of (\ref{gravcharges}) shows that, as one would expect, only those corresponding to the time translation Killing field, $\xi^{a}\partial_{a} = t^{a}\partial_{a} \equiv \partial/\partial t$ and the rotational Killing field $\xi^{a}\partial_{a} = \phi^{a}\partial_{a} \equiv \partial/\partial \phi$ are non-zero. They are given by:
\be Q_{t}[S] = \f{M}{(1+\f{a^2}{l^2})^2} \quad {\rm and} \quad
Q_{\phi}[S]  = - \f{Ma}{(1+\f{a^{2}}{l^{2}})^{2}} . \ee
These values are the $\Lambda >0$ counterparts of the Kerr-anti-de Sitter charges obtained in \cite{nd,sdrm}. Note that in the limit $\Lambda\to 0$ we have $\l\to \infty$ and the values of charges reduce to the standard Kerr charges.%
\footnote{The negative sign in front of the angular momentum charge is present also in the calculation at $\scri$ in asymptotically flat space-times if one uses the convention $l\cdot n =-1$ at $\scri$. The value of the mass is insensitive to this choice.}
However, in the presence of a cosmological constant, the metric has three parameters $M,a$ \emph{and} $l$,\, and the parameter $M$ is \emph{not} the charge generating time translation $\partial/\partial t$. It generates a rescaled time translation $\partial/\partial \bar{t}$ where $\bar{t} = (1+ (a/l)^{2})^{2}\, t$, where the rescaling varies from one space-time to another. In asymptotically Minkowski space-times, by `the' time translation, one means the one which is normalized to have unit norm at infinity with respect to the physical metric. In the presence of a cosmological constant, the norm of the `time translation' with respect to the physical metric diverges and so a canonical normalization for all space-times is not available. However, the first law of black hole thermodynamics restricts the freedom to change the normalization as one moves from one space-time to another \cite{afk}. Deruelle has shown that the (anti-de Sitter analogs of the) charges obtained here do satisfy the first law \cite{nd}.

\emph{Remark:}\\
In this sub-section we focused only on the time translation and three rotations because, as we discussed in section \ref{s4}, the completeness requirement in Definition 2 reduces $\Gds$ to this 4-dimensional group. However, because $q_{ab}$ is conformally flat, it admits 10 conformal Killing vectors $\xi^a$; it is just that the finite diffeomorphisms they generate fail to preserve the completeness condition. Since definitions of charges and fluxes do not refer to completeness, we can still use any of the 10 $\xi^{a}$ to define conserved charges $Q_{\xi}[S]$ and fluxes $\mathcal{F}_{\xi}^{\rm matt} [\Delta\scri]$. In the Kerr-de Sitter case, the additional 6 charges vanish on any $S$.

\subsubsection{Fluxes and balance laws}
\label{s5.3.4}

Let us begin with the dynamical collapse described by the Vaidya-de Sitter solution. The situation at $\scri^{+}$ is the same as in the Schwarzschild-de Sitter solution. However, since there is matter flux at $\scri^{-}$, the charge integrals are not conserved in the range $v_{1} \, < v\, <v_{2}$. (But because the source is a null fluid $\ub{T}$ vanishes on $\scri^{-}$ even in the dynamical region.) The leading term in the electric part of the Weyl tensor, $\mathcal{E}^{ab}$ has the same form as in the Schwarzschild-de Sitter space-time except that $M$ is not a constant but a function of $v$. Therefore, for any 2-sphere lying in the region $v\,<\, v_{1}$, we have $\mathcal{E}^{ab}=0$ and all charges vanish. In the dynamical region, the non-trivial balance law (\ref{balance1}) becomes relevant and the values of the `energy' charge integral increase in time in response to the matter flux $J^{a}$ flowing into the space-time across $\scri^{-}$. For $v \,>\, v_{2}$ the charge integral remains constant, and equals $M$. The charge integrals and the balance laws faithfully capture the energetics of the Vaidya solution because the underlying spherical symmetry implies that there are no gravitational waves, whence one knows that the energy flux is entirely due to matter. The overall situation parallels that in the Vaidya solutions in the $\Lambda =0$ case.

However, more generally, in the $\Lambda =0$ case the Bondi energy-momentum and angular momentum change also because of the leakage of gravitational waves across $\scri$. For example, for the charge corresponding to the time translation $\xi^{a} = \alpha n^{a}$ at $\scri$, the energy balance law reads
\be Q_{\xi}[S_{1}] -   Q_{\xi}[S_{2}] = \int_{\Delta \scri} \alpha\,\big[\frac{1}{32 \pi} \, |N_{ab}|^{2} +  \ub{T}_{ab} n^{a} n^{b} \big]\, \rmd^{3}V\, ,
\label{balance2}\ee
where the first term on the right hand side describes the flux of energy carried by gravitational waves. There is no analog of this term in (\ref{balance1}). Thus, if $\mathcal{B}^{ab} \, \hat{=} \, 0$, although the de Sitter charges are well-defined, there is no flux of de Sitter momentum due to gravitational waves even locally on $\scri$! In light of our discussion of the analogous condition in the $\Lambda =0$ case in section \ref{s5.1}, this could have been anticipated. For, when $\mathcal{B}^{ab} \, \hat{=} \, 0$ the Bondi news tensor $N_{ab}$ vanishes identically in the $\Lambda=0$ case and the gravitational contribution to the fluxes in the balance laws vanishes identically. That is, the balance law (\ref{balance1}) is the direct analog of the one in the $\Lambda=0$ case with additional restriction $\mathcal{B}^{ab} \, \hat{=} \, 0$. The parallel runs quite deep. For example, in the $\Lambda=0$ case the expression for the energy-momentum and angular momentum charge integrals is the same as the first term in (\ref{gravcharges}). The second term is absent simply because of the stronger fall-off of stress-energy tensor in the $\Lambda =0$ case.%
\footnote{Also, the overall multiplicative factor $l$ is absent simply because one normally chooses the conformal factor $\Omega \sim 1/r$ with dimensions of inverse length in the $\Lambda =0$ case while in the $\Lambda \not= 0$ case one chooses $\Omega \sim l/r$ which is dimensionless.} 
However, there is also a key difference. In the $\Lambda =0$ case, $\mathcal{B}^{ab}\, \hat{=} \,0$ implies that there is \emph{no} gravitational radiation at $\scri$ \emph{at all}; the equivalence class of connections $\{\mathring{D} \}$ is trivial. For $\Lambda >0$, this is not the case. Nothing prevents the electric part $\mathcal{E}^{ab}$ of the asymptotic Weyl tensor on $\scri$ from having a `transverse-traceless' piece in the decomposition of symmetric tensors into longitudinal, trace and transverse-traceless parts (that is often used in the initial value formulation of Einstein's equations). In \cite{abk2} we will see this feature in detail in the linearized approximation. In the $\Lambda =0$ case, by contrast, if $\mathcal{B}^{ab} \, \hat{=} \, 0$, the electric part $\mathcal{E}^{ab}$ is (again traceless but) longitudinal. 

Thus, the situation in the $\Lambda >0$ case is subtle. The condition $\mathcal{B}^{ab}\, \hat{=} \,0$ removes `half the radiative degrees of freedom' in the gravitational field and, in addition, the gravitational waves it does allow can not carry any of the de Sitter momenta across $\scri$. In this respect the situation is completely analogous to that for Yang-Mills fields in de Sitter space-time discussed in section \ref{s5.2}.\\

\emph{Remarks:}

1) As we noted earlier, our discussion in this sub-section parallels that in the $\Lambda <0$ case of \cite{aaam}. But in that case, the condition $\mathcal{B}^{ab}\, \hat{=} \,  0$ can be regarded as a reflective boundary condition \cite{swh}, an additional input that is needed to make the evolution well-defined because $\scri$ is time-like. The reflective nature of boundary conditions also explains why gravitational waves do not carry away energy-momentum or angular momentum across $\scri$. For $\Lambda >0$, on the other hand, $\mathcal{B}^{ab}\, \hat{=}\, 0$ is a genuine restriction because $\scri$ is space-like.

2) Abbott and Deser had analyzed the asymptotic structure of the gravitational field at spatial infinity already in the early 1980s \cite{ad}. They expanded the physical metric $g_{ab}$ around a de Sitter background $\bar{g}_{ab}$ as $g_{ab} = \bar{g}_{ab} + h_{ab}$ and defined gravitational charges using superpotentials constructed from $h_{ab}$ and its derivatives with respect to the background $\bar{g}_{ab}$. However given only $g_{ab}$, a priori,  there is some ambiguity in selecting the background de Sitter metric $\bar{g}_{ab}$. Therefore, analogs of supertranslation ambiguities can arise in the definitions of conserved charges (as briefly discussed in \cite{aaam}). 

This analysis was extended by Kastor and Traschen \cite{kt} to associate charges also with conformal isometries of the background de Sitter space-time and they were able to establish positivity of the charge associated with the conformal Killing field of $\bar{g}_{ab}$ which is timelike at $\scri$. Since this charge is associated with a conformal Killing field rather than Killing field, and since Einstein's equation is not conformally invariant, its physical interpretation is not transparent even in the linearized theory. Still, because it is positive, it could play a significant role, e.g., in various estimates used in geometrical analysis.

More recently, Kelly and Marolf \cite{km} have provided a Hamiltonian framework based on Cauchy data on space-like surfaces, analogous to the $\eta=$ const surfaces considered in section \ref{s3.5}, without reference to $\scri$. It is likely that their charges coincide with (\ref{gravcharges}) when our 2-spheres $S$ are chosen to lie in a neighborhood of $i^{o}$ within $\scri$ in which there is no matter flux. However, to firmly establish this result, one would have to understand the relation between the approach to $i^{o}$ along Cauchy surfaces and along $\scri$ more precisely. Finally, Chru\'{s}ciel, Jezierski and Kijowski have introduced the notion of a Hamiltonian mass for families of vacuum initial data sets with `ends of cylindrical type' \cite{clk}. To understand the relation between this mass and the conserved charges introduced here, one would first have to show that evolution of these data lead to space-times that are weakly asymptotically de Sitter in the sense of Definition 1.  

3) A natural strategy to relate frameworks based on $\scri$ to those based on (partial) Cauchy surfaces that go to spatial infinity, $i^{o}$, would be to introduce a 4-dimensional treatment of spatial infinity along the lines used in \cite{aarh,spatial-review,asymflat-others,aajr} for the $\Lambda=0$ case. In the Hamiltonian framework one works in the physical space-time and de Sitter charges arise as surface integrals on the 2-sphere boundaries of Cauchy surfaces at spatial infinity. Therefore, the treatment given in \cite{aajr} seems to be best suited for comparison because it attaches to the physical space-time a 3-dimensional boundary rather than a single point $i^{o}$. We examined this possibility in detail. However, it turns out that the structure for $\Lambda >0$ is so different that the basic ideas used in the $\Lambda=0$ construction do not generalize. In particular we could not endow the 3-dimensional `hyperboloid' at spatial infinity with a universal geometry as in the $\Lambda=0$ case. Therefore a space-time covariant treatment of spatial infinity remains an open problem.

4) In a Master's thesis, J\"{a}ger \cite{jager} has introduced charge integrals at $\scri$, following the procedure used in asymptotically anti-de Sitter space-times \cite{aaam}, just as we did in this section. However, as in \cite{ad,km}, a restriction was made to the source-free case. Therefore the charge integral did not have the second term in the expression (\ref{gravcharges}), nor the right side of our balance law (\ref{balance1}). Since the charges were absolutely conserved, the framework did not allow for situations analogous to the Vaidya collapse. Also the condition $\mathcal{B}^{ab} \, \hat{=} \, 0$  and the associated symmetry reduction carried out in section \ref{s4} was not considered, nor the relation to the $\Lambda=0$ case and Yang-Mills fields discussed in this section. 

5) In the literature inspired by the AdS/CFT correspondence, to define gravitational charges one generally introduces infinite counter terms to handle the blow up of the metric components at infinity in the commonly used charts \cite{adscft}. By contrast, in the framework used here, everything is manifestly finite and no subtractions are necessary because the formulation is in terms of the Weyl curvature which remains smooth (and in fact vanishes!) at $\scri$.

\section{Discussion}
\label{s6}

We began in section \ref{s2} by summarizing and slightly extending the standard constructions that have been used in the literature to discuss asymptotically de Sitter space-times. In section \ref{s3} we presented several examples to illustrate the finer differences one finds in different physical situations, particularly in the topology of $\scri$. These examples also brought out the differences from the more familiar asymptotically Minkowski space-times. 

In section \ref{s4} we discussed asymptotic symmetries in detail. We found that in general asymptotically de Sitter space-times, the asymptotic symmetry group $\G$ is just $\Diff$ whence we cannot repeat the procedures used in the $\Lambda=0$ case to extract physics from the asymptotic behavior of the gravitational field at $\scri$. This  surprising outcome has not been appreciated in the literature in part because a stronger boundary condition is often introduced that requires the intrinsic 3-metric at $\scri$ to be conformally flat. Then, the symmetry group $\G$ reduces to the 10-dimensional de Sitter group $\Gds$ if the topology of $\scri$ is $\mathbb{S}^{3}$. However, in physically interesting space-times the topology is generally different. In the cosmological context it is often $\mathbb{R}^{3}$ and for isolated gravitating systems it is $\mathbb{S}^{2} \times \mathbb{R}$. We showed that completeness requirement on $\scri$ reduces the symmetry group further to a 7-dimensional group in the $\mathbb{R}^{3}$ case and to a 4-dimensional group of a `time' translation and 3 rotations in the $\mathbb{S}^{3}\times \mathbb{R}$ case. 

In section \ref{s5} we showed that given these asymptotic symmetries we can introduce conserved charges. In the Kerr-de Sitter space-time, and in the dynamical Vaidya-de Sitter solution depicting the simplest black hole formation via gravitational collapse, the conserved charges can be evaluated explicitly and provide the physically expected mass and angular momentum. However, we also showed that even in fully dynamical space-times where one expects gravitational waves near $\scri$, the charges are absolutely conserved if there is no matter flux across $\scri$. {Thus, one is apparently led to conclude that \emph{gravitational waves do not carry energy or angular momentum across} $\scri$ if $\Lambda >0$! However, this severe limitation comes from the stronger conformal flatness condition.} We showed that it is equivalent to asking that the magnetic part $\mathcal{B}^{ab}$ of the leading order Weyl curvature at $\scri$ must vanish. Thus, while the stronger condition seems attractive from symmetry considerations, it removes \emph{by fiat} half the degrees of freedom. These results are surprising because they imply that we do not yet have a strategy to extract physics of gravitational waves, and more generally, properties of isolated gravitating systems, in full general relativity with positive $\Lambda$. To better understand this limitation, we discussed in some detail the implications of the condition $\mathcal{B}^{ab} \,= \,0$ in the $\Lambda =0$ case and of the analogous condition ${B}^{a}_{i}\, \hat{=}\, 0$ on Yang-Mills fields on de Sitter space-time. We found that in both cases, the condition imposes unreasonably severe restrictions on permissible fields and constrains all the local fluxes of energy, momentum and angular momentum across $\scri$ to vanish identically.

As pointed out in section \ref{s1}, this leads to a quandary: if we drop the requirement of conformal flatness, the structure at $\scri$ is too weak to extract physics and if we keep it, we rule out the examples that are of primary interest to gravitational wave science and quantum gravity and of significant interest to geometric analysis.

There are further issues of prime physical interest in these three areas that are difficult to investigate using the currently available constructions. We will present one example in each area. At the interface of general relativity and geometric analysis, positive energy theorems are not only major landmarks but also serve as invaluable tools if $\Lambda$ vanishes. However, in the $\Lambda >0$ case, we cannot even speak of de Sitter momentum unless we impose the stronger $\mathcal{B}^{ab} \, \hat{=}\, 0$ condition, which eliminates the possibility of accounting for energy, momentum and angular momentum loss due to gravitational waves. Furthermore, now  all symmetry vector fields are space-like near $\scri$ because $\scri$ itself is space-like. Therefore, one cannot hope to prove positive energy theorems either at $\scri$ or at spatial infinity $i^{o}$.%
\footnote{A related issue is the black hole uniqueness theorem in 4 dimensions which has also remained wide open for rotating black holes when $\Lambda=0$ even though (unlike in the $\Lambda <0$ case) the horizon is guaranteed to have a 2-sphere topology.} 
Indeed for test fields in de Sitter space-time, energy can be arbitrarily negative even when all the local energy conditions are satisfied simply because the `time translation' vector fields are all space-like near and on $\scri$. On the other hand, there is a time-translation Killing field which is time-like in the static patch of de Sitter space-time, whence the flux of energy of test fields across the future and the past cosmological horizons is guaranteed to be positive (see Fig. \ref{ds-mink}). Can one perhaps extend this idea to full non-linear general relativity and obtain positive energy theorems? 

 \begin{figure}[]
  \begin{center}
    \includegraphics[width=3in,height=2.5in,angle=0]{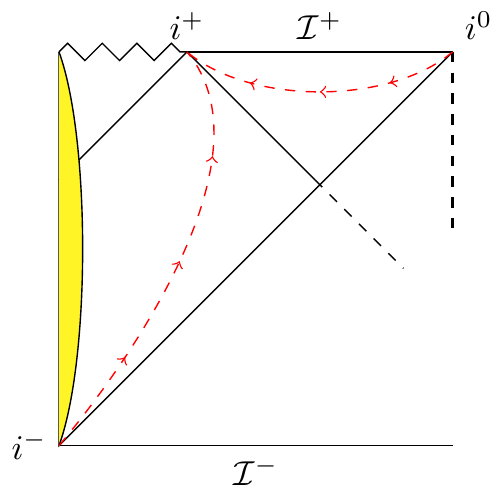}
    
\caption{Conformal diagram depicting a spherical collapse. Shaded (yellow) region corresponds to the collapsing spherical star. The  dashed (red) lines with arrows represent integral curves of the `static' Killing field. Note that the space-time is incomplete to the right unless we add another collapsing star.} 
\label{collapse}
\end{center}
\end{figure}
Next, let us consider gravitational collapse leading to the formation of a black hole. In the case of a Schwarzschild-de Sitter black hole, the well-known Kruskal conformal diagram of the $\Lambda =0$ case is replaced by Fig. \ref{sch}. We now have cosmological horizons and, furthermore, we have to carry out an identification if we want to avoid an infinite number of black hole and white hole regions. Because of this identification, Cauchy surfaces now have topology $\mathbb{S}^{2}\times \mathbb{S}^{1}$ (rather than $\mathbb{S}^{2}\times \mathbb{R}$). Now consider gravitational collapse leading to the formation of a Schwarzschild-de Sitter black hole depicted in Fig. \ref{collapse}. Again, the space-time diagram continues indefinitely to the right but now because the collapsing region is dynamical, a natural identification is not possible unless one adds another collapsing star to the right. While one can do this in the spherical case \cite{shapiro}, it seems difficult to envisage the analogous extension for a generic collapse. In any case, physically, one generally needs to consider the collapse of a single star, e.g., to study the Hawking effect. Then we are led to Fig. \ref{collapse} which shows that it will not suffice to specify the incoming state at $\scri^{-}$ alone, since additional information can flow in from the time-like dashed line on the right of the diagram. And it is difficult to know what the appropriate additional data would be to capture the idea that the total incoming state be vacuum. Once the back reaction is included, further ambiguities arise at $\scri^{+}$.

Finally, consider the problem of black hole coalescence. Detailed investigations we are aware of have been carried out in the $\Lambda =0$ case. In that case, the $\scri$ framework is rich and, in particular, it enables one to calculate the 3-momentum that is radiated away across $\scri^{+}$ (in the center of mass frame defined by the ADM 4-momentum). To compensate for this loss, the final black hole recoils, giving rise to the celebrated `black hole kicks' of astrophysical interest that have been studied in detail in numerical relativity \cite{kicks}. If $\Lambda >0$, gravitational waves do not carry away any energy or momentum or angular momentum across $\scri^{+}$. Does this then mean that in our real universe with a positive $\Lambda$ there are no kicks? More generally, what are the implications to our actual universe with $\Lambda>0$  of all the beautiful simulations in the $\Lambda=0$ case that have provided us with detailed estimates of energy and angular momentum loss across $\scri^{+}$ in binary coalescences? These are important issues that must be addressed now, since we are at the threshold of the golden era of gravitational wave science through the global networks of advanced detectors. 

The goal of the subsequent papers in this series is to construct a theoretical platform to address these issues and also develop systematic techniques to estimate errors one makes by setting $\Lambda =0$ from the start. In view of the smallness of the observed value of $\Lambda$, we expect these effects to be small. But we need a clean framework to say with confidence how small they are even when the sources are at cosmological separations and gravitational waves have been traveling distances of the order of a giga-parsec before reaching us. In the linearized approximation, these calculations can be performed with the current theoretical tools (as, e.g., in \cite{bicaketal, abk2}). However, keeping in mind the initial confusion on the physical reality of gravitational waves in \emph{full, non-linear} general relativity, we need the analog of the Bondi framework in full general relativity to satisfactorily address these issues. Considering the vast literature in the $\Lambda <0$ case, it is rather puzzling that very basic \emph{physical} issues remain in the $\Lambda >0$ case, in spite of its direct relevance to the universe we inhabit.   

\section*{Acknowledgment}

We would like to thank Piotr Chru\'{s}ciel, Hans Ringstr\"om and David 
Robinson for correspondence; Eugenio Bianchi, Luc Blanchet, Alejandro Corichi, Jo\~{a}o Costa, Josh Goldberg and Peter Saulson for discussions; and  Ji\v{r}\'{i}  Bi\v{c}\'{a}k for providing some early references that we were unaware of. This work was supported in part by the NSF grant PHY-1205388, the Eberly research funds of Penn State and a Frymoyer Fellowship to AK.

\appendix
\section{Asymptotically Minkowski space-times}
\label{sa1}

In this Appendix we prove a technical result that was used as an intermediate step in section \ref{s5.1} to show that, in asymptotically Minkowski space-times, if $\mathcal{B}^{ab} \, \hat{=} \, 0$ then the Bondi news tensor must vanish, i.e., $N_{ab} \, \hat{=} \,  0$. 

In section \ref{s5.1} we saw that $\mathcal{B}^{ab} \, \hat{=} \, 0$ implies that $N_{ab}$ is a lift to $\scri$ of a tensor field $\ub{N}_{ab}$ on the 2-sphere $S$ of generators of $\scri$, satisfying:
\be \ub{N}_{[ab]} =0, \quad \ub{N}_{ab}\, \ub{q}^{ab} =0,\quad {\rm and} \quad \ub{D}_{[a}\, \ub{N}_{b]c} =0  \label{barN} \ee
where $\ub{q}_{ab}$ is the metric on $S$ induced by the metric $q_{ab}$ on $\scri$ and $\ub{D}$ its derivative operator. While this result holds in any conformal frame, it is easiest to extract its consequences by working in a Bondi conformal frame where $\ub{q}_{ab}$ is the unit 2-sphere metric. In the rest of this Appendix, we make this assumption.

Let us embed $(S, \ub{q}_{ab})$ as the unit 2-sphere in $\mathbb{R}^{3}$ and denote the Cartesian coordinates in $\mathbb{R}^{3}$ by $x^{i}$. Then, the projections $\ub{C}^{a} := \ub{q}^{a}{}_{b} \Co^{b}$ of constant vector fields $\Co^{b}$ on $\mathbb{R}^{3}$ provide us with `pure' conformal Killing fields on $(S, \ub{q}_{ab})$:
\be \ub{C}_{a} = \ub{D}_{a} \Co^{b}x_{b} = \ub{q}_{a}^{c}\, \partial_{c} (\Co_{b}x^{b}), \quad {\rm and}\quad \ub{D}_{a}\ub{C}_{b} = - (\Co_{c}x^{c}) \,\ub{q}_{ab} . \ee 
(Restrictions of functions $\Co_{c}x^{c}$ to $S$ are linear combinations of the first three spherical harmonics $Y_{1,m}$.) It is easy to verify that $\ub{D}_{[a}\, (\ub{N}_{b]c} \ub{C}^{c})\, = 0$. Therefore $\ub{N}_{bc}\ub{C}^{c}$ is a gradient $\ub{D}_{b} h$ on $S$ and we can eliminate the freedom of adding a constant to its potential $h$ by requiring $\oint_{S}\, h\, d^{2}V =0$. 

Thus, $\ub{N}_{ab}$ provides us with a linear map from the 3-dimensional space of `pure conformal Killing fields' $\ub{C}^{a}$ on $(S, \ub{q}_{ab})$ to the space of functions $h$ on $S$ (satisfying  $\oint_{S}\, h d^{2}V =0$). But there is also a natural isomorphism between the vector space of these $\ub{C}^{a}$ and the vector space of their conformal Killing data \cite{aaam2} at any point $p$ on $S$, 
\be \big(\ub{C}^{a},\,\, \ub{D}_{[a} \ub{C}_{b]},\,\, \ub{D}_{a}\ub{C}^{a},\,\, \ub{D}^{a}\ub{D}_{b}\ub{C}^{b}\big)_{p} \, = \, (\ub{q}^{ab}\ub{D}_b \Co \cdot x,\,\, 0,\,\, (\Co\cdot x), \,\, \ub{q}^{ab}\ub{D}_b (\Co \cdot x) )_{p}\, .\ee
Since the conformal Killing data is determined by the values of  $\ub{C}^{a}$ and $(\Co\cdot x)$ in the tangent space of every point $p$ on $S$, we have a number $\ub{f}$ and a vector $\ub{V}_{a}$ such that the function $h$ determined by $\ub{N}_{ab}$ is given at $p$ by:
\be h = [\ub{q}^{ac}\ub{D}_{c}(\Co\cdot x)]\, \ub{V}_{a} \, +\, 
 [\Co \cdot x]\, \ub{f} . \ee
Now, using the relation $\ub{N}_{ab}\, \ub{C}^{b} = \ub{D}_{a} h$ and the fact that $\ub{N}_{ab}$ is tangential to $S$ and therefore satisfies $\ub{N}_{ab} x^{b} =0$, one concludes  $\ub{V}_{a} = \ub{D}_{a} \ub{f}$ and hence
\be \quad \ub{N}_{ab} = \ub{D}_{a}\ub{D}_{b} \ub{f} + \ub{f} \ub{q}_{ab} . \label{keyeq} \ee
This is the key consequence of the first and the third equations in (\ref{barN}). Finally we use the second equation. The fact that $\ub{N}_{ab}$ is trace-free implies that $f$ must satisfy $\ub{D}^{2} \ub{f} + 2 \ub{f} =0$ whence it is necessarily of the form $\ub{f} = \Co \cdot x$ for some $\Co$. Substituting this back in (\ref{keyeq}) implies $\bar{N}_{ab} \, \hat{=} \,  0$. Since $N_{ab}$ is the pull-back to $\scri$ of $\bar{N}_{ab}$, we conclude $N_{ab} \, \hat{=} \,  0$.\\

\emph{Remark:}\\
Since $S$ has the topology of $\mathbb{S}^{2}$, the conclusion $\ub{N}_{ab} =0$ follows already from $\ub{N}_{[ab]} =0$ and $\ub{D}^{a}\ub{N}_{ab} =0$ on $S$ which are implied by (\ref{barN}) \cite{aaam3}. The proof we presented here is of interest because of the intermediate result (\ref{keyeq}) which follows only form the first and the third equation in (\ref{barN}), and \emph{more importantly} holds also on surfaces of constant curvature with higher dimensions and non-positive definite signature. (However, if the signature is not positive definite, the trace-free property of $\ub{N}_{ab}$ does not then imply $\ub{N}_{ab} =0$.) The higher dimensional result is used in the analysis of spatial infinity \cite{aarh,spatial-review,asymflat-others}.

\end{document}